\documentclass{aa}  

\usepackage{graphicx}
\usepackage{txfonts}
\usepackage{courier}
\usepackage[version=3]{mhchem}
\usepackage{hyperref}
\usepackage{array,booktabs,ragged2e}
\usepackage{amsmath}
\makeatletter
\@ifundefined{linenumbers}{}{
  \let\linenumbers\relax
  \let\nolinenumbers\relax

  \let\do@linenumbers\relax
  
}
\makeatother
\makeatletter
\renewcommand*\aa@manuscriptname{}     
\renewcommand*\aa@journalname{}        
\makeatother
\begin{document}

   \title{Chemical modeling of aminoketene, ethanolamine, and glycine production in interstellar ices}
   \titlerunning{Chemical modeling of aminoketene}

   \author{Sydney A. Willis,
          \inst{1}
          Serge A. Krasnokutski,
          \inst{2}
          Nathaniel J. Morin
          \inst{1}
          \and
          Robin T. Garrod
          \inst{1,3}
          }
   \institute{Department of Chemistry, University of Virginia, Charlottesville, VA 22904 \\
              \email{swillis@virginia.edu}
             \and
             Laboratory Astrophysics Group of the Max Planck Institute for Astronomy, Friedrich Schiller University Jena, D-07743 Jena, Germany \\
             \email{sergiy.krasnokutskiy@uni-jena.de}
             \and
             Department of Astronomy, University of Virginia, Charlottesville, VA 22904 }
 
\abstract
{Icy interstellar dust grains are a source of complex organic molecule (COM) production, although the formation mechanisms of these molecules are debated. Laboratory experiments show that atomic carbon deposited onto interstellar ice analogs can readily react with solid-phase ammonia to form the \ce{CHNH2} radical, a possible precursor to COMs, including aminoketene (\ce{NH2CHCO}).}
{We used astrochemical kinetics models to explore the role of the reaction of atomic C with ammonia as well as the subsequent reaction with CO in the formation of aminoketene and other COMs, including ethanolamine (\ce{NH2CH2CH2OH}) and glycine (\ce{NH2CH2COOH}).}
{We applied the three-phase chemical model MAGICKAL to hot molecular core conditions from the cold-collapse through to the hot-core stage. The chemical network was extended to include \ce{NH2CHCO} and a range of associated gas-phase, grain-surface, and bulk-ice products and reactions. We also implemented a model approximating conditions in a shocked cloud, including sputtering of the ice mantles. }
{Aminoketene is formed on grains at low temperatures ($\sim$10~K) with a peak solid-phase abundance of $\sim$2$\times$10$^{-10}$~n$_{\mathrm{H}}$. Its formation is driven by nondiffusive reactions, in particular the Eley-Rideal reaction of C with surface NH$_3$, followed by immediate reaction with CO.
Surface hydrogenation of aminoketene produces ethanolamine with a significant abundance of $\sim$8$\times$10$^{-8}$ n$_{\mathrm{H}}$. 
In the gas-phase, although ethanolamine reaches a modest abundance peak immediately following its desorption from grains under hot-core conditions, it is destroyed more rapidly due to its high proton affinity. Molecular survival is much higher in the shocked regions, where these species seem most likely to be detected.
Glycine abundances are modestly enhanced by the new chemistry.}
{Aminoketene is produced efficiently on simulated interstellar grain surfaces, acting subsequently as an important precursor to more complex organics, including ethanolamine and glycine. Ion-molecule gas-phase destruction of amine-bearing COMs is less efficient in (weakly) shocked lower-density regions, in contrast to hot cores, enhancing their abundances and lifetimes.}

   \maketitle
\section{Introduction}
    \label{Intro}

In astrochemistry, complex organic molecules (COMs) are defined as carbon-bearing species with six or more atoms \citep{H+VD2009}. As of 2023, more than 300 COMs have been detected in the interstellar medium (ISM), including in cold molecular clouds, hot cores, hot corinos, and protoplanetary disks\footnote{https://cdms.astro.uni-koeln.de/classic/molecules} \citep{CDMS}. For many years, the prevailing theory for the formation of COMs primarily involved gas-phase ion-molecule chemistry driven by the thermal desorption of simpler organic species such as \ce{CH3OH} \citep{Charnley1995,Viti99,Jorgensen2020}. 
While current research suggests that neutral-neutral reactions in the gas phase may indeed be important for COM production \citep[e.g.][]{Barone2015,Skouteris2018,Skouteris2019}, ion-molecule reactions are found not to be uniformly efficient in producing COMs, partly due to the high probability of breakup associated with dissociative electronic recombination \citep{Geppert2005,Hamberg10}. Gas-phase production of the ubiquitous interstellar molecule methyl formate (HCOOCH$_3$) appears particularly inefficient \citep{Horn2004,Garrod2006}.

Over the past two decades, much attention has been paid to the possibility of grain-surface and ice chemistry to account for the COMs found abundantly in hot cores and corinos. Chemical kinetics models of such sources \citep{Garrod2006,Garrod2008} have indicated that surface diffusion of radicals driven by elevated temperatures, on the order of 20--40~K, could allow the fragments of photodissociated simple molecules to combine into larger structures. Photochemical processing of simple grain-surface ices continues to indicate the importance of photon-driven COM production \citep[e.g.,][]{Schutte1993,Oberg2009,MartinDom2020,Ishibashi2021}. However, although heating of the grains can enhance production by promoting the recombination of surface and ice radicals \citep{Butscher2016}, observational evidence indicates that COMs can be present in significant quantities even at the very low temperatures achieved in prestellar cores \citep[e.g.][]{Bacmann2012,JS2016}. More recently, \citet{Rocha2024} have reported the detection of solid-phase COMs toward high-mass and low-mass protostellar sources based on JWST data.

Laboratory experiments have indicated that oxygen-bearing organic species in particular can be formed on very cold surfaces by the deposition of H atoms onto surface ices containing CO and/or related species such as formaldehyde \citep[e.g.][]{Fedoseev2015,Fedoseev2017,Chuang2016}. Radicals produced via hydrogenation or H-abstraction can then rapidly recombine to form COMs without the necessity for diffusion. Other combinations of simple species allow even the simplest amino acid, glycine (NH$_2$CH$_2$COOH), to form in a similar manner without the diffusion of large radicals or by photoprocessing \citep{Ioppolo2021}. \citet{J&G2020} and \citet{G22} presented new rate formulations to incorporate nondiffusive reaction mechanisms into the gas-grain chemical models and networks necessary to simulate COM chemistry in both cold and hot cores; they indeed found that many COMs, including methyl formate, glycine, and others, can form via an array of processes from low to high temperatures, with many COMs forming at the very early stages of cold cloud collapse.

Recent experimental and theoretical work has concentrated on the influence of atomic carbon adsorption on icy surfaces. Adsorption of atomic C directly onto surface water leads to the production of formaldehyde with a low activation energy barrier \citep{Hickson2016,Molpeceres2021,Potapov2021}, circumventing the usual expectation that surface formaldehyde production requires prior formation of CO. Condensation of C atoms together with other abundant ISM molecules, namely H$_2$, H$_2$O, and CO, results in the formation of refractory organic molecules with a large oxygen content \citep{Krasnokutski2017}. 

\begin{figure}
    \label{energy diagram}
    \centering
    \includegraphics[width=1\linewidth]{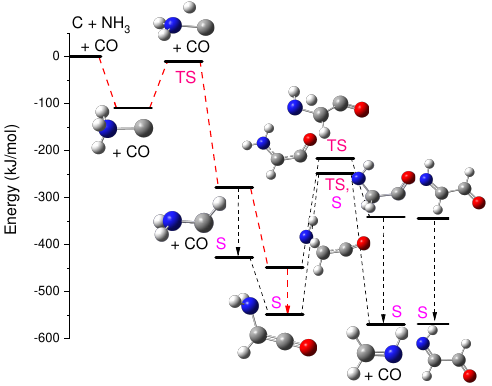}
    \caption{Energy level diagram of the C + NH$_3$ + CO reaction. The reaction proceeds from left to right. It starts with a triplet state; the intersystem crossings to the singlet states are marked with dashed arrows. S and TS stand for singlet and transition states. Red lines indicate the most probably pathway.}
    \label{lvl1}
\end{figure}

The reaction between atomic C and ammonia has been investigated using two different experimental approaches. In the first, individual reactant pairs were examined at T = 0.37 K using a calorimetry technique \citep{Henning2019}, as reported by \citet{Krasnokutski2020}. Within 1 millisecond, which is the duration of the experiment, products such as CHNH$_2$ or CH$_2$NH in electronically excited triplet states were detected. A subsequent study employed a more conventional cryogenic molecular ice technique to explore the interaction of C atoms with pure NH$_3$ ice or NH$_3$:CO ice mixtures at 10 K \citep{Krasnokutski2022}. For all ices, the emergence of infrared absorption bands corresponding to CH stretching modes immediately after carbon atom deposition provided clear evidence of H transfer from nitrogen to carbon. This observation supports the formation of CHNH$_2$ or CH$_2$NH species and is consistent with the findings of the first study. 
Considering the presence of the energy well in the reaction pathway associated with the C$\bullet$NH$_3$ product as shown in Figure~\ref{lvl1}, the only explanation that is consistent with all the experimental and theoretical results is the H transfer before relaxation. This is also consistent with expectations, given that intramolecular proton transfer occurs on a femtosecond timescale \citep{Chou01}.
The formed products may then go on to react with other surface species to form larger organic and even biological molecules, including peptides \citep{Krasnokutski21,Krasnokutski2022}. In this context, the reactivity of CHNH$_2$ is essential. This is in contrast to the recently proposed chemical reaction network \citep{Molpeceres24}, in which the condensation of C atoms on the surface of ice was suggested to result in the formation of a relatively weakly bound complex, C$\bullet$NH$_3$. 

The formation of CHNH$_2$ in particular, which can rapidly react with CO, would appear to open up different reaction pathways than those associated with C$\bullet$NH$_3$. The alternative pathway, rapid production of CH$_2$NH, would also lead to different reactions than might be expected if C$\bullet$NH$_3$ were retained as the product.

Of particular astrochemical interest for this ice chemistry are so-called prebiotic complex organics: COMs that may be important in the eventual formation of life \citep{Zeng2021}. Prebiotic COMs tend to be rich in nitrogen, in the form of amine, nitrile, or amide functional groups \citep{Zeng2023}. Primary amines (-\ce{NH2}) are an important component of the structure of peptides and polypeptides (i.e., proteins), which are composed of amino acids.
Generally, peptides are considered to be formed directly from the condensation of amino acids, but this process has a high energy barrier and therefore usually occurs at temperatures higher than 300 K \citep{Kitadai2018,Steele2020,Iqubal2017}.

Like other complex species, these prebiotic COMs have been identified in a variety of sources, although \ce{NH2}-bearing COMs are typically found only toward high-mass star-forming regions, such as Sagittarius (Sgr) B2(N) \citep{Belloche2017}. More recently, some \ce{NH2}-bearing species have been identified in the quiescent molecular cloud G+0.693-0.027 \citep{Zeng2018}, hereafter referred to as G+0.693; this Galactic Center cloud appears to have undergone shocks, releasing the ice mantles via sputtering \citep{Rivilla2022}. Table~\ref{detected species} shows a list of amine group-bearing species detected toward Sgr B2(N) and G+0.693 and fractional abundances with respect to H$_2$.

Considering the divergent physical conditions and evolution of G+0.693 versus high-mass star-forming sources such as Sgr B2(N2), the formation of the \ce{NH2}-bearing COMs found in these sources may be more dependent on their earlier physical conditions \citep{Jorgensen2020} or simply on their common potential for the formation of COM-rich ices during those earlier periods.

   \begin{table}
      \caption[]{Amine-bearing species detected in the ISM. }
         \label{detected species}
         \centering
         \begin{tabular}{llr}
            \hline
            \noalign{\smallskip}
            Species & Source & Fractional  \\
                    &        & Abundance  \\
            \noalign{\smallskip}
            \hline
            \noalign{\smallskip}
            \ce{NH2CHO} & Sgr B2(N1S)$^a$ & (2.0 $\pm$ 0.8) $\times$ 10$^{-6}$ \\
            & Sgr B2(N2)$^b$ & (2.5 $\pm$ 0.7) $\times$ 10$^{-6}$ \\
            & G+0.693-0.027$^c$ & (9.3 $\pm$ 2.0) $\times$ 10$^{-10}$ \\
            \noalign{\smallskip}
            \ce{CH3(CO)NH2} & Sgr B2(N1S)$^a$ & (2.9 $\pm$ 0.8) $\times$ 10$^{-7}$ \\
            & Sgr B2(N2)$^a$ & (9.9 $\pm$ 0.8) $\times$ 10$^{-8}$ \\
            & G+0.693-0.027$^c$ & (8.5 $\pm$ 0.1) $\times$ 10$^{-10}$ \\
            \noalign{\smallskip}
            \ce{CH3NH2} & Sgr B2(N1S)$^d$ & (9.9 $\pm$ 0.8) $\times$ 10$^{-7}$ \\
            & G+0.693-0.027$^e$ & (2.2 $\pm$ 0.5) $\times$ 10$^{-8}$ \\
            \noalign{\smallskip}
            \ce{NH2(CO)NH2} & Sgr B2(N1S)$^a$ & (1.9 $\pm$ 0.8) $\times$ 10$^{-8}$ \\
            & G+0.693-0.027$^c$ & (1.0 $\pm$ 0.1) $\times$ 10$^{-10}$ \\
            \noalign{\smallskip}
            \ce{NH2CH2CN} & Sgr B2(N1S)$^a$ & (7.7 $\pm$ 0.8) $\times$ 10$^{-8}$ \\
            \noalign{\smallskip}
            
            \noalign{\smallskip}
            \hline
            \noalign{\smallskip}

         \end{tabular}
         \tablefoot{Fractional abundances are given with respect to \ce{H2}. References: a)\citet{Belloche2019}; b)\citet{Bonfand2019};  c)\citet{Zeng2023}; d)\citet{Zeng2018}; e)\citet{Melosso2020}}
   \end{table}

Given the evidence outlined above for low-temperature, nondiffusive grain-surface COM-formation mechanisms, such chemistry may play an important role in the formation of \ce{NH2}-bearing prebiotic COMs. Based on their experiments, \citet{Krasnokutski2022} and \citet{Krasnokutski2024} have proposed the barrierless surface formation of \ce{NH2CHCO}, followed by its polymerization, as an energetically favorable mechanism for the formation of peptides in interstellar and other astrophysical conditions. In addition to polymerization, \ce{NH2CHCO} would be expected to undergo repetitive hydrogenation reactions with atomic H under interstellar conditions, potentially leading to other relevant prebiotic COMs such as ethanolamine (\ce{NH2CH2CH2OH}). The availability of the OH radical in a water-rich interstellar ice could similarly promote the formation of glycine (\ce{NH2CH2COOH}), via OH addition to an intermediate radical derived from \ce{NH2CHCO} hydrogenation. 

The grain-surface pathway to the formation of \ce{NH2CHCO} involves two successive reactions (reactions (1a) and (2) in Table~\ref{reactions}):
\begin{gather}
    \label{CHNH2}
    \ce{C} + \ce{NH3} \rightarrow \ce{CHNH2} \notag \\
    \label{AMK}
    \ce{CHNH2} + \ce{CO} \rightarrow \ce{NH2CHCO}. \notag
\end{gather}

The energy level diagram for these reactions is shown in Figure~\ref{lvl1}. The ground state of the C atoms defines the initial triplet state of the system. The reaction starts with the formation of the weakly bound complex C$\bullet$NH$_3$ and is followed by the transfer of the hydrogen from nitrogen to carbon. As discussed above, this reaction is extremely fast, so that the weakly bound complex does not thermalize, allowing the transfer of H over the energy barrier. Although singlet states are the most energy favorable for products of the C + NH$_3$ reaction, the products remain in the triplet state for at least one millisecond \citep{Krasnokutski2020}, allowing their complete thermalization in the triplet state. In this state, the \ce{NH2CHCO} isomer is the most energy favorable. After its thermalization and intersystem crossing to the singlet state, the reaction is frozen, since propagation of the reaction to the right along the energy level diagram with the formation of the most energetically favorable singlet states, namely NHCHCHO or CH$_2$NH + CO, requires the overcoming of relatively high energy barriers, which cannot take place at low temperatures. 

Here, using astrochemical kinetics models, we examine the efficiency of this reaction scheme on the production of nitrogen-bearing COMs, both in star-forming cores and under conditions appropriate to a shocked cloud.
The chemical network is extended to consider the formation and destruction of aminoketene (\ce{NH2CHCO}), ethanolamine (\ce{NH2CH2CH2OH}), and glycine (\ce{NH2CH2COOH}), and a range of related species.

Section~\ref{methods} describes the new reactions incorporated into the chemical network, as well as a new mechanism for the incorporation of shocks. Section~\ref{sec:results} presents the results of the hot-core chemical model with the updated network, in addition to the results of the shock simulation. Section~\ref{disc} compares the chemical model results to observational values and examines the astrochemical implications of the work, followed by conclusions in Section~\ref{concl}. 

\section{Methods}
\label{methods}

\subsection{Astrochemical modeling}
\label{modeling overview}

To explore the chemistry of the \ce{NH2CHCO} that might be formed in the ices of interstellar dust grains, we constructed a new set of grain-surface and gas-phase chemical reactions related to \ce{NH2CHCO} and associated species based on the core findings of \citet{Krasnokutski2020,Krasnokutski2022}. We added these reactions to the existing network used in the astrochemical kinetics model MAGICKAL \citep{G2013,G22,GH2023}, which was then applied to astrophysical conditions, with a primary focus on the chemistry of hot cores. Additional setups were run to approximate the shock conditions in G+0.693.

MAGICKAL utilizes a three-phase treatment, which includes gas, grain-surface, and bulk-ice mantle chemistry to simulate the coupled gas-grain chemistry of interstellar clouds and cores, and has been used frequently to model the chemistry of hot cores. It uses a rate equation-based method and a modified-rate method for the surface chemistry, which alters reaction rates to approximate stochastic behavior when deemed necessary \citep{G2008}. The grain-surface chemistry framework considers diffusive grain chemistry as well as a range of nondiffusive processes consisting of so-called three-body (3-B) reactions, photodissociation-induced (PDI) reactions, and Eley-Rideal (E-R) reactions; see \citet{G22} for details of the implementation. All grain-surface reactions included in the network are allowed to proceed through each of these meeting mechanisms. Bulk-ice chemistry includes 3-B and PDI reactions, while diffusive chemistry may also occur for reactions involving H or H$_2$, which are the only species allowed to undergo bulk diffusion.

The gas-phase chemical network includes various ion-molecule and neutral-neutral reactions, photodissociation processes by both external UV or cosmic ray-induced UV, radiative association, and electronic recombination and dissociation reactions. Proton-transfer reactions involving gas-phase molecular ions are also included, following \citet{GH2023}.

The hot-core physical model employs two stages, with the first stage corresponding to the cold, free-fall collapse and the second stage corresponding to the warm-up of the hot core at fixed gas density. The free-fall collapse stage begins at a density $n_{\mathrm{H}} = 3000$~cm$^{-3}$, visual extinction $A_V = 3$, gas temperature $T_{\mathrm{gas}} = 10$~K, and dust temperature $T_{\mathrm{dust}} \simeq 14.7$~K, which is determined from a calculation based on $A_V$ \citep{GP11}. The collapse continues over $\sim$10$^{6}$~yr to a final density $2 \times 10^{8}$~cm$^{-3}$, visual extinction of 500 mag, and $T_{\mathrm{dust}} \simeq 8$~K. $T_{\mathrm{gas}}$ remains constant at 10~K during the collapse. During the warm-up stage, $n_{\mathrm{H}}$ is held constant, while $T_{\mathrm{gas}}$ and $T_{\mathrm{dust}}$ increase to 400~K over $2.85 \times 10^{5}$~yr, based on $T_{\mathrm{dust}}$ reaching 200~K at $2 \times 10^{5}$~yr. This warm-up scale corresponds to the medium timescale in previous publications. The cosmic ray ionization rate (CRIR) is constant throughout, taking the canonical value $1.3 \times 10^{-17}$~s$^{-1}$ \citep{DW1993}.

Although our main focus is hot-core chemistry, we also run chemical models under shock conditions; see Sec.~\ref{sec:shock} for a detailed description of the method.

\subsection{Addition of aminoketene and products to the network}
\label{sec:aminoketene}

The chemical network used in these simulations is based on the network of \citet{GH2023}, labeled as M5 in that work. However, the present network also includes propanol chemistry, as described by \citet{Belloche2022}, and dimethylamine chemistry, as described by \citet{Muller2023}. The network has been expanded in this work to include reactions involving \ce{NH2CHCO} and selected products, including ethanolamine (\ce{NH2CH2CH2OH}), glycine (\ce{NH2CH2COOH}), and glycinal (\ce{NH2CH2CHO}); the latter two species, along with associated radicals, were already present in the earlier networks \citep{G2013} and thus have several radical-addition processes by which they can be formed aside from those added here.

The addition of new complex species in the network is handled in the same way as in past models \citep[e.g.,][]{Belloche2022}. The newly added neutral species and their binding energies, which have been extrapolated from species with similar functional groups, are listed in Table~\ref{species}. 

Table~\ref{reactions} shows the new grain-surface and ice-mantle reactions leading to aminoketene, as well as a selection of important related reactions added to the network. These reactions can occur as a result of diffusive or nondiffusive chemistry (E-R for surface only, 3-B, and PDI). The reactions include both addition and abstraction of H from certain species; both processes (where included) are assumed to be barrierless in the case of radicals, whereas nonzero activation energy barriers are assumed for closed-shell species. Activation energies are obtained by quantum chemical calculations. The calculations were performed using the hybrid functional B3LYP with the 6-311+G (d, p) basis set implemented in the Gaussian16 package \citep{G16}. To obtain the reactions' energy barriers, the scan along the reactive coordinates was performed with complete optimization of reactive complex geometries. B3LYP offers a good balance between accuracy and computing time required for scanning the reaction coordinates with molecular structure optimization for all reactions.

The new network also includes a number of new grain-surface reactions that are included for consistency with the existing network, following the approach of \cite{G2013} (see their Tables 2 and 3). Such reactions include H-atom addition to all new radicals, as well as H-atom transfer reactions between new radicals and existing radicals such as HCO and \ce{CH2OH} to produce two closed-shell species (reactions 17--18). Most new radical addition (single-product) reactions added to the network are listed in Table~\ref{reactions}. Other reactions added for completeness include barrierless H-abstraction from new radicals by reactive radicals such as OH and NH$_2$ (reactions 19 and 20), and barrier-mediated H-abstraction by those radicals from the new closed-shell species (reactions 21 and 22). 

Reaction numbers referred to in this paper correspond specifically to the numbering system used in Table~\ref{reactions}. Reactions that were already present in the network, or those that are of minor importance, are generally not shown in the table.

\begin{table}
    \caption[]{Binding energies and proton affinities for relevant neutral species.}
        \label{species}
        \centering
        \begin{tabular}{lcr}
        \hline
        \noalign{\smallskip}
        {Species} & $E_{\mathrm{des}}$ (K) & PA (kJ mol$^{-1}$)\\
        \noalign{\smallskip}
        \hline
        \noalign{\smallskip}
        {\ce{NH2CHCO}} & 5 381 & n/a \\
        {\ce{NH2CH2CH2OH}} & 10 017 & 930.3$^a$\\
        {\ce{NH2CH2$\dot{\mathrm{C}}$HOH}} & 7 267 & 856.0$^b$ \\
        {\ce{NH2CH2CH2$\dot{\mathrm{O}}$}} & 9 356 & 856.0$^b$ \\
        {\ce{NH2$\dot{\mathrm{C}}$HCHO}} & 6 042 & n/a \\
        {\ce{$\dot{\mathrm{C}}$HNH2}} & 4 231 & n/a \\
        {\ce{NH2CH2CHO}$^*$} & 6 606 & 856.0$^b$ \\
        {\ce{NH2CH2COOH}$^*$} & 10 126 & 886.5$^a$ \\
        \noalign{\smallskip}
        \hline
        \end{tabular}
        \tablefoot{Radicals are marked with a dot at the radical site. $^*$ indicates a species that was already present in the network, which we note here for completeness. The values for binding energies are extrapolated from similar species or those with similar functional groups in the network, and are based on the work of \citet{G+WW+H2008}. The values for proton affinities are taken from multiple sources a) \citet{H+L1998} and b) \citet{GH2023}.}
\end{table}

The new chemistry is based around the production of \ce{NH2CHCO} from reactions (1a) and (2) occurring on the grains, either on the surface or in the bulk ice. We note that, due to the femtosecond conversion timescale of C$\bullet$NH$_3$ to either \ce{CHNH2} or \ce{CH2NH}, C$\bullet$NH$_3$ does not need to be explicitly included in the network. The \ce{CHNH2} radical may also be formed from \ce{CH2NH2} via H-abstraction by another radical. At low temperatures ($<$20~K), reactions (1) and (2) will proceed predominantly through nondiffusive surface reactions (E-R, 3-B, or PDI). At higher temperatures ($>$20~K), they may may also proceed via surface diffusion. In the bulk ice, these reactions are allowed to occur only through the nondiffusive 3-B and PDI processes. Because atomic C is crucial to the process, reactions on the surface, as opposed to the bulk ice, will tend to dominate, as much of this carbon will be adsorbed from the gas phase.

After the formation of \ce{NH2CHCO}, glycine (\ce{NH2CH2COOH}) can form via the further addition of a hydrogen atom (3a), followed by the addition of OH to the resulting radical (10a):
\begin{gather}
    \label{glycine}
    \ce{NH2CH2CO} + \ce{OH} \rightarrow \ce{NH2CH2COOH}, \notag
\end{gather}

\noindent which at low temperatures will be driven by nondiffusive processes. Reaction (10a) was already present in the network, but an alternative branch, reaction (10b) was added here, assuming a statistical branching ratio. Reaction (3a) can proceed through Langmuir-Hinshelwood diffusion of H on the surface and in the bulk ice, or through nondiffusive chemistry. Although aminoketene is a closed-shell species, it is found experimentally to be very reactive, and we calculate only a small barrier to reaction (3a). The reaction between these species may also proceed through channel (3b), although with a more substantial barrier, to produce the alternative radical form, \ce{NH2CHCHO}.

Following hydrogenation via reactions (3a) or (3b), further hydrogenation can occur to form glycinal and ethanolamine via reactions (4--8): 
\begin{gather}
    \label{glycinal}
    \ce{NH2CHCO} + \ce{2H} \rightarrow \ce{NH2CH2CHO} \notag \\
    \label{EtA}
    \ce{NH2CH2CHO} + \ce{2H} \rightarrow \ce{NH2CH2CH2OH}. \notag
\end{gather}
These steps also include H-abstraction channels (see Table~\ref{reactions}). Of the four possible reactions of H with \ce{NH2CH2CHO} included in our network, listed as reactions (6a--d), abstraction back to the radical \ce{NH2CH2CO} has the lowest barrier, but a fraction will still be hydrogenated all the way to ethanolamine. Reactions (9a,b) are plausible H-abstraction reactions involving ethanolamine and atomic H. We note that, for the key barrier-mediated reactions (3a-b) and (6a-d), at the approximately 10~K temperatures at which ice production is operative, the reactions proceed substantially faster than the competing process of H diffusion away from the reactant. Thus, in spite of the barriers, the probability of reaction through one of the branches of reactions (3) or (6) is close to 100\% for every meeting of an H-atom with the reactant species.

    \begin{table*}
      \caption[]{Selection of grain-surface and ice-mantle reactions added to the MAGICKAL chemical network.}
         \label{reactions}
         \centering
         \begin{tabular}{llclclclr}
            \hline
            \noalign{\smallskip}
            Reaction & & & & & & & & $E_{\mathrm{A}}$ (K) \\
            \noalign{\smallskip}
            \hline
            \noalign{\smallskip}
            1a & \ce{C} & + & \ce{NH3} & $\rightarrow$ & \ce{CHNH2} &  & & 0 \\
            1b & \ce{C} & + & \ce{NH3} & $\rightarrow$ & \ce{CH2NH} & \smallskip & & 0 \\

            2 & \ce{CHNH2} & + & \ce{CO} & $\rightarrow$ & \ce{NH2CHCO} & \smallskip & & 0 \\

            3a & \ce{H} & + & \ce{NH2CHCO} & $\rightarrow$ & \ce{NH2CH2CO} &  & & 61 \\
            3b & \ce{H} & + & \ce{NH2CHCO} & $\rightarrow$ & \ce{NH2CHCHO} & \smallskip & & 750 \\

            4 & \ce{H} & + & \ce{NH2CH2CO} & $\rightarrow$ & \ce{NH2CH2CHO} & \smallskip & & 0 \\

            5a & \ce{H} & + & \ce{NH2CHCHO} & $\rightarrow$ & \ce{NH2CH2CHO} &  & & 0 \\
            5b & \ce{H} & + & \ce{NH2CHCHO} & $\rightarrow$ & \ce{NH2CHCO} & + & \ce{H2} \smallskip &  545 \\

            6a & \ce{H} & + & \ce{NH2CH2CHO} & $\rightarrow$ & \ce{NH2CH2CH2O} &  & & 1 717 \\
            6b & \ce{H} & + & \ce{NH2CH2CHO} & $\rightarrow$ & \ce{NH2CH2CHOH} &  & & 2 612 \\
            6c & \ce{H} & + & \ce{NH2CH2CHO} & $\rightarrow$ & \ce{NH2CH2CO} & + & \ce{H2} & 895 \\
            6d & \ce{H} & + & \ce{NH2CH2CHO} & $\rightarrow$ & \ce{NH2CHCHO} & + & \ce{H2} \smallskip & 1 625 \\
            
            7a & \ce{H} & + & \ce{NH2CH2CH2O} & $\rightarrow$ & \ce{NH2CH2CH2OH} & & & 0 \\
            7b & \ce{H} & + & \ce{NH2CH2CH2O} & $\rightarrow$ & \ce{NH2CH2CHO} & + & \ce{H2} \smallskip & 0 \\
            
            8 & \ce{H} & + & \ce{NH2CH2CHOH} & $\rightarrow$ & \ce{NH2CH2CH2OH} & \smallskip & & 0 \\
            
            9a & \ce{H} & + & \ce{NH2CH2CH2OH} & $\rightarrow$ & \ce{NH2CH2CH2} & + & \ce{H2O} & 7 400 \\
            9b & \ce{H} & + & \ce{NH2CH2CH2OH} & $\rightarrow$ & \ce{NH2CH2CHOH} & + & \ce{H2} \smallskip & 2 134 \\

            10a$^*$ & \ce{OH} & + & \ce{NH2CH2CO} & $\rightarrow$ & \ce{NH2CH2COOH} & & & 0 \\
            10b & \ce{OH} & + & \ce{NH2CH2CO} & $\rightarrow$ & \ce{NH2CHCO} & + & \ce{H2} \smallskip & 0 \\

            11a & \ce{H} & + & \ce{CHNH2} & $\rightarrow$ & \ce{CH2NH} & + & H & 0 \\
            11b & \ce{H} & + & \ce{CHNH2} & $\rightarrow$ & \ce{CHNH} & + & \ce{H2} & 0 \\
            11c & \ce{H} & + & \ce{CHNH2} & $\rightarrow$ & \ce{CH2NH2} & & \smallskip & 0 \\

            12 & \ce{H} & + & \ce{CH2NH2} & $\rightarrow$ & \ce{CH3NH2} & & \smallskip & 0 \\

            13 & \ce{CH} & + & \ce{NH2} & $\rightarrow$ & \ce{CHNH2} & & \smallskip & 0 \\

            14 & \ce{O} & + & \ce{CHNH2} & $\rightarrow$ & \ce{NH2CHO} & & \smallskip & 0 \\

            15 & \ce{H2} & + & \ce{CHNH2} & $\rightarrow$ & \ce{CH3NH2} & & \smallskip & 4 010$\dag$ \\
            
            16 & \ce{CH4} & + & \ce{CHNH2} & $\rightarrow$ & \ce{CH3CH2NH2} & & \smallskip & 12 200$\ddag$ \\

            17a & \ce{HCO} & + & \ce{NH2CH2CH2O} & $\rightarrow$ & \ce{NH2CH2CH2OH} & + & \ce{CO} & 0 \\
            17b & \ce{HCO} & + & \ce{NH2CH2CH2O} & $\rightarrow$ & \ce{NH2CH2CHO} & + & \ce{H2CO} \smallskip & 0 \\

            18a & \ce{CH2OH} & + & \ce{NH2CH2CH2O} & $\rightarrow$ & \ce{NH2CH2CH2OH} & + & \ce{H2CO} & 0 \\
            18b & \ce{CH2OH} & + & \ce{NH2CH2CH2O} & $\rightarrow$ & \ce{NH2CH2CHO} & + & \ce{CH3OH} \smallskip & 0 \\

            19 & \ce{OH} & + & \ce{NH2CH2CH2O} & $\rightarrow$ & \ce{NH2CH2CHO} & + & \ce{H2O} \smallskip & 0 \\

            20 & \ce{NH2} & + & \ce{NH2CH2CH2O} & $\rightarrow$ & \ce{NH2CH2CHO} & + & \ce{NH3} \smallskip & 0 \\

            21 & \ce{OH} & + & \ce{NH2CH2CHO} & $\rightarrow$ & \ce{NH2CHCHO} & + & \ce{H2O} \smallskip & 597 \\

            22 & \ce{NH2} & + & \ce{NH2CH2CHO} & $\rightarrow$ & \ce{NH2CHCHO} & + & \ce{NH3} \smallskip & 3 390 \\
           
            \noalign{\smallskip}
            \hline
         \end{tabular}
         \tablefoot{For new reactions, the activation energies were calculated as part of this work (described in Section~\ref{Intro}). In the case of diffusive reactions with \ce{CHNH2}, more than one activation energy was available, depending on the spin state of the molecule. In each case, the lower energy was included in the model. $\dag$ indicates a triplet state, and $\ddag$ indicates a singlet state. $^*$ indicates a reaction that was already present in the network, which we note here for completeness.}
   \end{table*}

Several grain-surface and bulk ice reactions reactions (11a--c) between \ce{CHNH2} and atomic H are included, one of which leads to conversion to \ce{CH2NH}. All three branches are assumed to proceed without barriers and are assigned statistical branching. Similar reactions involving \ce{CH2NH} already exist in the network (mediated by activation energy barriers) and are not shown in the table. Reaction (13) is also included in the network, whereby \ce{CHNH2} is formed by the direct addition of CH and NH$_2$.

Three other reactions involving \ce{CHNH2} were included in the network (14--16), corresponding to three other plausible reaction partners, atomic O, H$_2$ and CH$_4$. Our calculations indicate that the latter two have substantial activation energy barriers.

Besides the grain-surface reactions and processes, a basic gas-phase chemistry is added to the network for all new species, as per past publications. Although the larger COMs, with their high binding energies, will tend to remain on the grains at low temperatures, the high temperatures associated with hot cores will allow them to be thermally desorbed. Once in the gas phase, proton transfer followed by dissociative electronic recombination is typically the main destruction mechanism, via reactions with protonated molecular species such as \ce{H3+}, \ce{H3O+}, or \ce{HCO+}. Which reaction is dominant depends on the molecular content of the gas, which is related to the temperature regime via the desorption of, for example, H$_2$O and CO. The resulting protonated species will recombine with an electron, which is assumed to break up the molecule in 95\% of collisions.

In the case where a chemical species has a proton affinity less than that of ammonia ($<$853.6~kJ~mol$^{-1}$), the collision of its protonated form with ammonia allows it to transfer the proton to ammonia and thus be returned to its nonprotonated form, avoiding electronic recombination and therefore enhancing its gas-phase lifetime \citep{TWC2016}. Species with proton affinities greater than that of ammonia will experience rapid gas-phase destruction by protonation or recombination, as they are unable to shed the proton by collisions with abundant ammonia, while collision with protonated ammonia will lead to their protonation. Cosmic rays are also a source of destruction for gas-phase species, either by ionization or direct dissociation reactions. Rates for all of these processes are calculated following methods used in past models \citep{GH2023}.

\subsection{Aminoketene production parameters}
\label{sec:production}

To test \ce{NH2CHCO} production in hot cores, we run a selection of models in which the parameters controlling the production of \ce{NH2CHCO} on grains are varied. Table~\ref{input} indicates the specified parameters for each model setup. The {\tt Basic} model indicates a standard setup for the MAGICKAL model, with the parameters as described in Section~\ref{modeling overview} and the new chemistry involving \ce{NH2CHCO} production as described in Section~\ref{sec:aminoketene}. An {\tt Old} model, without the new chemical species or reactions, is also included as a reference for species that were already present in the chemical network (e.g.,~\ce{NH2CH2CHO} and \ce{NH2CH2COOH}).
   \begin{table*}
      \caption[]{Input parameters for tested models.}
         \label{input}
         \centering
         \begin{tabular}{llccc}
            \hline
            \noalign{\smallskip}
            Model ID & Description     &  BR & (Final) Gas Density & CR ionization rate \\
            \multicolumn{1}{l}{} &  &   & (cm$^{-3}$) & (s$^{-1}$) \\
            \noalign{\smallskip}
            \hline
            \noalign{\smallskip}
            {\tt Old} & Hot core; Model M5 of \citet{GH2023} & n/a & 2 $\times$ 10$^{8}$ & 1.3 $\times$ 10$^{-17}$ \\
            {\tt Basic} & Hot core; new \ce{NH2CHCO} chemistry & 50/50 & 2 $\times$ 10$^{8}$ & 1.3 $\times$ 10$^{-17}$ \\
            {\tt BR--high} & Hot core; \ce{CHNH2}-favored & 100/0 & 2 $\times$ 10$^{8}$ & 1.3 $\times$ 10$^{-17}$  \\
            {\tt BR--low} & Hot core; \ce{CH2NH}-favored & 10/90 & 2 $\times$ 10$^{8}$ & 1.3 $\times$ 10$^{-17}$  \\
            {\tt BR--off} & Hot core; Reactions 1 and 2 excluded & n/a & 2 $\times$ 10$^{8}$ & 1.3 $\times$ 10$^{-17}$  \\
            {\tt S--LCR} & Shock model; low cosmic-ray ionization rate & 50/50 & 2 $\times$ 10$^{4}$  & 1.3 $\times$ 10$^{-17}$  \\
            {\tt S--MCR} & Shock model; medium cosmic-ray ionization rate & 50/50 & 2 $\times$ 10$^{4}$  & 1.3 $\times$ 10$^{-16}$  \\
            {\tt S--HCR} & Shock model; high cosmic-ray ionization rate & 50/50 & 2 $\times$ 10$^{4}$ & 1.3 $\times$ 10$^{-15}$  \\
            {\tt S--VHCR} & Shock model; very high cosmic-ray ionization rate & 50/50 & 2 $\times$ 10$^{4}$ & 1.3 $\times$ 10$^{-14}$  \\
            \noalign{\smallskip}
            \hline
         \end{tabular}
   \end{table*}

In addition to the formation of \ce{CHNH2} via reaction (1a), the reaction of atomic carbon and \ce{NH3} can also form \ce{CH2NH}, via reaction (1b). The \ce{CH2NH} product has a considerable barrier against reaction with CO, and thus its formation will not lead to the production of aminoketene. The energy levels corresponding to \ce{CHNH2} and \ce{CH2NH}, the products of reactions (1a,b) are very close in the triplet states, which are defined by the ground state of the C atom and where the separation between isomers takes place. The \ce{CH2NH} product is about 1500~K lower in energy \citep{Krasnokutski2020}. These products are formed due to H transfer from nitrogen to carbon. A second H transfer leading to the \ce{CH2NH} molecule is associated with a significant energy barrier of about 18900~K. Therefore, the second H transfer will occur only if it is faster than the thermalization of the \ce{CHNH2} molecule. Experimental evidence has shown that the first H is transferred before the system is thermalized, which suggests that the rate of the second H transfer could also be high \citep{Krasnokutski2020}. Additionally, the second H transfer can take place only in the case if there is no nearby CO molecule, while in the opposite case, the barrierless reaction with CO and the formation of \ce{NH2CHCO} will dominate, and the formation of \ce{CH2NH} does not allow the formation of peptides, which contradict the experimental observations \citep{Krasnokutski2022}. Finally, we explored the possibility that the second H transfer can occur without any restriction. In that case, the ratio of the occupancies of both levels corresponding to the \ce{CHNH2} and \ce{CH2NH} molecules can be determined using the Boltzmann factor. For this calculation, the temperature is defined by the height of the energy barrier (18900~K) that separates these levels, resulting in an almost equal amount of both molecules with the ratio being about 1.08 in favor of the most energetically favorable \ce{CH2NH} molecule.

To accommodate these uncertainties, we utilized three different models: the {\tt Basic}, {\tt BR--high}, and {\tt BR--low} models. In the {\tt Basic} model, the efficiency of each reaction is considered to be 50\%. {\tt BR--high} and {\tt BR--low} are included to test the outcome if the efficiency of reactions 1 and 2 are more or less than what we assume them to be in the {\tt Basic} setup. In the case of \ce{CHNH2} being favored, we do not allow \ce{CH2NH} to be produced at all from the reaction of C and \ce{NH3}. In addition, the {\tt BR--off} model is included to test the importance of reactions (1) and (2) in \ce{NH2CHCO} production. In this model, the efficiencies of reactions (1a,b) and (2) are set to 0\%, but \ce{NH2CHCO} can still be produced through a series of hydrogen-abstractions (16 and 17), assuming there is sufficient energy for the reaction barriers to be overcome.

\subsection{Shock model}\label{sec:shock}

With the addition of \ce{NH2CH2CH2OH}, the MAGICKAL chemical network now contains six amine-bearing species that have been detected in shocked molecular cloud G+0.693. Thus, we have also adapted MAGICKAL to include a shock treatment to approximate these physical conditions. Much of the shock treatment follows that of \citet{JS2008}, who modeled the steady-state physical parameters of a C-shock using a parametric model, including a sputtering treatment. 

The method of \citet{JS2008} approximates the neutral and ion velocities as functions of the spatial variable and the shock velocity. The neutral gas density is determined by conservation of mass, and is a function of the shock velocity, the neutral gas velocity, and the initial gas density. The temperatures of the neutral and ionized components of the gas are also approximated, based on the peak shock temperature provided by fitting data from \citet{Draine83}. The method further relates the spatial position with respect to the shock front with the time passed for a parcel of gas moving through the shock, allowing the physical conditions to be imposed on a time-dependent chemical model simulating such a gas parcel. We have based our calculations of the gas temperatures, gas density, and gas velocities explicitly on the subroutines provided in the online version of the chemical code UCLCHEM\footnote{https://uclchem.github.io/}.

Perhaps most crucial for the chemical model is the determination of the dust temperature and the rate of sputtering, which are required to determine the rates of loss of ice material into the gas phase. \citet{JS2008} based their sputtering rates on the drift velocity, i.e.,~the velocity difference between ions and neutrals. However, as noted by \citet{Miura2017}, the grain heating and sputtering rates depend on the relative rates of the neutral gas and the dust grains themselves. Here, we follow Miura et al.~by integrating their Eq.~7 (and associated formulae) for the dust velocity; we take the aforementioned approximated neutral gas velocity and temperature as time-dependent input values. We simultaneously integrate their Eq.~9 to get the dust temperature. All calculations assume the same grain radius $a = 0.1~\mu$m that is used in the chemical model. With the time/position-dependent dust velocity in hand, the difference in dust and neutral gas velocities can be used to calculate the ice sputtering rates during the chemical simulations.

All of these time-dependent physical conditions are fed into a two-stage chemical model. In stage 1, the pre-shock stage, the cloud undergoes a collapse in a similar way as in the hot core models. The final chemical abundances from stage 1 are then passed on to stage 2, when the cloud undergoes shock conditions as described above. The time-dependent physical conditions used in stage 2 are all calculated in advance, allowing the chemical calculations to be kept entirely separate from the shock treatment. The time-dependent neutral gas temperature from the shock model is used as the (only) gas temperature in the chemical model during stage 2, i.e.,~the calculated ion temperature does not influence the chemistry.

The sputtering rates of the ice mantles used in stage 2 are calculated within MAGICKAL, following the method of \citet{JS2008}, which is based on formulations by \citet{DS79} and \citet{Draine95}. This calculation method produces a rate of overall ice loss from the grains that is dependent on the instantaneous shock physical conditions; in our model, this means the gas temperature and the difference in dust and neutral gas velocities. Calculations of the overall sputtering rate also assume that the ice mantle is composed of water ice, while all gas-phase neutral species are allowed to act as sputtering agents. For the purposes of the sputtering calculations, the abundances of those species are set at the beginning of the shock process and held fixed until the shock conditions have ended. With the overall ice sputtering rate calculated at each moment, the loss rates of individual ice species in our model are based on the overall sputtering rate multiplied by each species' instantaneous proportional abundances in the outer ice layer in the chemical model. Because only low-velocity shocks are considered here, grain-core sputtering can be ignored.

\begin{figure}
    \centering
    \includegraphics[width=0.495\textwidth]{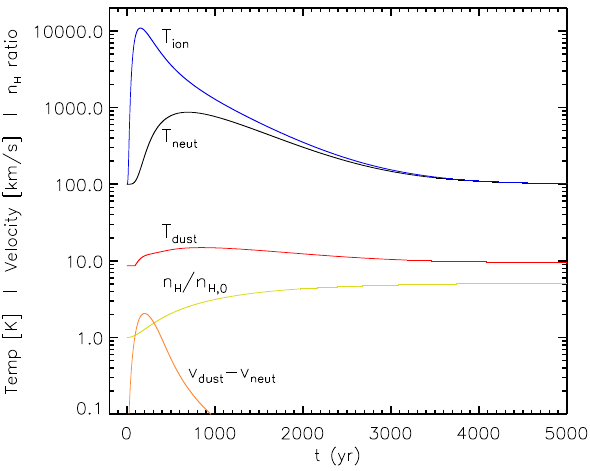}
    \caption{Physical parameters produced by the parametric shock model, assuming a 20 km s$^{-1}$ shock speed, initial gas density $n_{\mathrm{H}} = 2 \times 10^4$~cm$^{-3}$, and a grain size $a = 0.1$~$\mu$m. The difference between the speeds of the neutral gas and the dust grains is important in determining the rate of ice sputtering in the chemical model. The shock treatment is based on the methods of \citet{JS2008} and \citet{Miura2017}.}
    \label{shock_physics}
\end{figure}

Following \citet{Rivilla2022}, we adopt shock input parameters specific to G+0.693 of $v_s=20$~km~s$^{-1}$ and cloud density $n_\mathrm{H} = 2 \times 10^4$~cm$^{-3}$. The stage 1 models therefore terminate their collapse at this density, reaching a minimum dust temperature of 8.6~K. The gas temperature is held at 100~K throughout stage 1 to represent the average kinetic temperature of G+0.693. Stage 2 of the shock models corresponds to the actual passage of the shock, and does not include a gradual warm-up of the gas and dust temperature as in the hot-core models; instead, there is a rapid warm-up during the shock to a temperature of $\sim$870 K. Due to the low speed of the shock, and the low gas density that produces only modest thermal coupling, the dust temperature rises only to a peak value of $\sim$13~K. The peak relative gas-dust velocity is a little over 2~km~s$^{-1}$. Figure~\ref{shock_physics} shows the behavior of the various temperatures, the gas density, and the differential between dust and neutral gas velocities in the model, as a function of linear time. Beyond the 5000~yr timescale shown, the physical conditions remain static. Below a value of $v_{\mathrm{dust}}-v_{\mathrm{neut}} \simeq 0.1$~km~s$^{-1}$, the rate of ice sputtering becomes negligible.

According to \citet{Goto2014}, the cosmic-ray ionization rate (CRIR) can be above 100 times greater than the local Galactic rate \citep[$\sim$10$^{-17}$~s$^{-1}$;][]{Padovani2009}. Additionally, \citet{Riv2022} have suggested that the CRIR can be as high as 1,000 times the local Galactic rate, based on observations and modeling of \ce{PO+}. Therefore, we have run four shock models, {\tt S--LCR}, {\tt S--MCR}, {\tt S--HCR}, and {\tt S--VHCR} that have varying cosmic-ray ionization rates, but which all otherwise employ the same physical input conditions.

\section{Results}\label{sec:results}

\begin{figure*}
    \centering
    \includegraphics[width=0.435\textwidth]{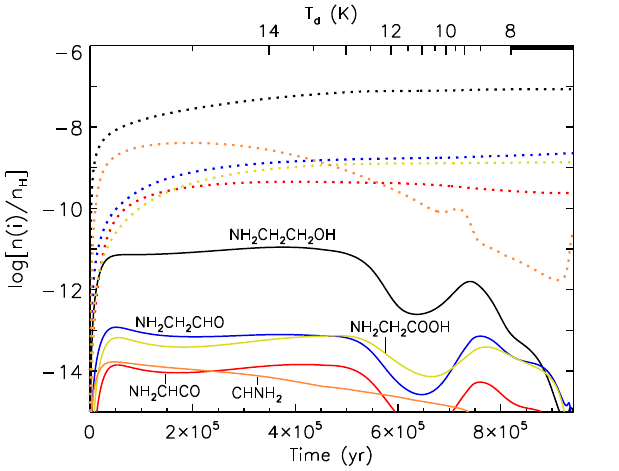}
    \includegraphics[width=0.435\textwidth]{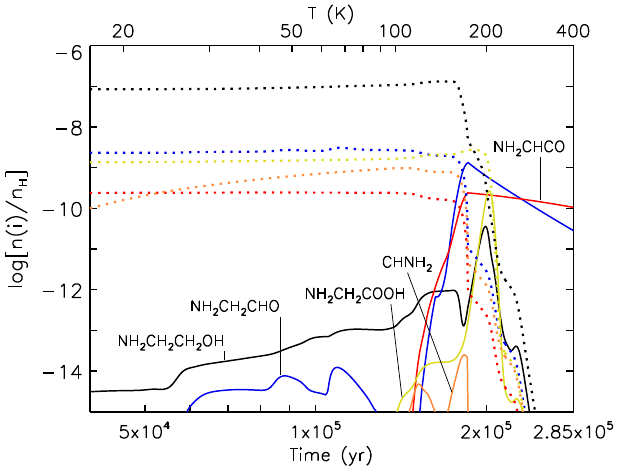}
    \caption{Fractional abundances of \ce{NH2CHCO} and relevant products with respect to total hydrogen. The results correspond to the {\tt Basic} setup, with the collapse stage shown on the left and the warm-up stage shown on the right. Solid lines indicate gas-phase abundances; dotted lines of the same color indicate the species on the grain (surface and bulk ice combined). The thicker bar in the upper axis of the left panel indicates that the dust temperature remains at 8~K once that value is reached.}
    \label{basic}
\end{figure*}

Tables~\ref{all_st1_output} and~\ref{all_st2_output} present chemical abundances of \ce{NH2CHCO} and related products from each of the hot-core models. Table~\ref{all_st1_output} shows the modeled solid-phase fractional abundances with respect to total hydrogen of the selected species at the end of Stage 1 (collapse), while Table~\ref{all_st2_output} shows the peak gas-phase fractional abundances of the same selected species during Stage 2 (warm-up). Figure~\ref{basic} shows time-dependent molecular abundances for the {\tt Basic} models in the collapse and warm-up stages. The behavior of each of the hot-core models is described below; shock model results are described in Sec.~\ref{results:shock}.

   \begin{table*}
      \caption[]{Final solid-phase fractional abundances of \ce{NH2CHCO} and relevant products during the collapse (stage 1).}
         \label{all_st1_output}
         \centering
         \begin{tabular}{lccccc}
            \hline
            \noalign{\smallskip}
            Model ID & \ce{NH2CHCO} & \ce{NH2CH2CH2OH} & \ce{NH2CH2COOH} & \ce{NH2CH2CHO} & \ce{CHNH2} \\
            \noalign{\smallskip}
            \hline
            \noalign{\smallskip}
            {\tt Old} & N/A & N/A & 8.17 $\times$ 10$^{-10}$ & 8.04 $\times$ 10$^{-10}$ & N/A \\
            {\tt Basic} & 2.29 $\times$ 10$^{-10}$ & 8.36 $\times$ 10$^{-8}$ & 1.32 $\times$ 10$^{-9}$ & 2.18 $\times$ 10$^{-9}$ & 2.85 $\times$ 10$^{-11}$ \\
            {\tt BR--high} & 2.07 $\times$ 10$^{-10}$ & 1.29 $\times$ 10$^{-7}$ & 1.44 $\times$ 10$^{-9}$ & 2.14 $\times$ 10$^{-9}$ & 2.33 $\times$ 10$^{-11}$ \\
            {\tt BR--low} & 2.40 $\times$ 10$^{-10}$ & 4.82 $\times$ 10$^{-8}$ & 1.35 $\times$ 10$^{-9}$ & 2.26 $\times$ 10$^{-9}$ & 3.14 $\times$ 10$^{-11}$ \\
            {\tt BR--off} & 1.34 $\times$ 10$^{-11}$ & 1.23 $\times$ 10$^{-8}$ & 5.71 $\times$ 10$^{-10}$ & 3.80 $\times$ 10$^{-10}$ & 6.86 $\times$ 10$^{-12}$ \\
            {\tt S--LCR} & 6.25 $\times$ 10$^{-10}$ & 2.21 $\times$ 10$^{-7}$ & 4.27 $\times$ 10$^{-9}$ & 4.89 $\times$ 10$^{-9}$ & 2.32 $\times$ 10$^{-11}$ \\
            {\tt S--MCR} & 1.17 $\times$ 10$^{-10}$ & 1.11 $\times$ 10$^{-7}$ & 3.33 $\times$ 10$^{-9}$ & 2.24 $\times$ 10$^{-9}$ & 1.29 $\times$ 10$^{-11}$ \\
            {\tt S--HCR} & 2.59 $\times$ 10$^{-11}$ & 6.88 $\times$ 10$^{-8}$ & 3.36 $\times$ 10$^{-9}$ & 5.88 $\times$ 10$^{-10}$ & 1.08 $\times$ 10$^{-11}$ \\
            {\tt S--VHCR} & 1.37 $\times$ 10$^{-12}$ & 5.08 $\times$ 10$^{-9}$ & 2.28 $\times$ 10$^{-10}$ & 2.60 $\times$ 10$^{-11}$ & 2.45 $\times$ 10$^{-11}$ \\
            \noalign{\smallskip}
            \hline
         \end{tabular}
         \tablefoot{Abundances are given with respect to total hydrogen (n$_H$).}
   \end{table*}

   \begin{table*}
      \caption[]{Peak gas-phase fractional abundances (n$_H$) of \ce{NH2CHCO} and relevant products during the warm-up or shock (stage 2).}
         \label{all_st2_output}
         \centering
         \begin{tabular}{lccccc}
            \hline
            \noalign{\smallskip}
            Model ID & \ce{NH2CHCO} & \ce{NH2CH2CH2OH} & \ce{NH2CH2COOH} & \ce{NH2CH2CHO} & \ce{CHNH2} \\
            \noalign{\smallskip}
            \hline
            \noalign{\smallskip}
            {\tt Old} & N/A & N/A & 1.68 $\times$ 10$^{-10}$ & 4.66 $\times$ 10$^{-10}$ & N/A \\
            {\tt Basic} & 2.30 $\times$ 10$^{-10}$ & 3.50 $\times$ 10$^{-11}$ & 2.60 $\times$ 10$^{-10}$ & 1.26 $\times$ 10$^{-9}$ & 2.37 $\times$ 10$^{-14}$ \\
            {\tt BR--high} & 2.14 $\times$ 10$^{-10}$ & 4.90 $\times$ 10$^{-11}$ & 2.79 $\times$ 10$^{-10}$ & 1.27 $\times$ 10$^{-9}$ & 2.14 $\times$ 10$^{-14}$ \\
            {\tt BR--low} & 2.40 $\times$ 10$^{-10}$ & 2.28 $\times$ 10$^{-11}$ & 2.66 $\times$ 10$^{-10}$ & 1.28 $\times$ 10$^{-9}$ & 2.40 $\times$ 10$^{-14}$ \\
            {\tt BR--off} & 2.02 $\times$ 10$^{-11}$ & 7.35 $\times$ 10$^{-12}$ & 1.21 $\times$ 10$^{-10}$ & 2.02 $\times$ 10$^{-10}$ & 4.89 $\times$ 10$^{-15}$ \\
            {\tt S--LCR} & 1.65 $\times$ 10$^{-12}$ & 4.73 $\times$ 10$^{-10}$ & 1.08 $\times$ 10$^{-11}$ & 1.31 $\times$ 10$^{-11}$ & 1.00 $\times$ 10$^{-14}$ \\
            {\tt S--MCR} & 8.85 $\times$ 10$^{-13}$ & 8.82 $\times$ 10$^{-10}$ & 1.99 $\times$ 10$^{-11}$ & 1.11 $\times$ 10$^{-11}$ & 5.85 $\times$ 10$^{-15}$ \\
            {\tt S--HCR} & 4.05 $\times$ 10$^{-13}$ & 1.54 $\times$ 10$^{-9}$ & 4.15 $\times$ 10$^{-11}$ & 3.72 $\times$ 10$^{-12}$ & 8.10 $\times$ 10$^{-15}$ \\
            {\tt S--VHCR} & 3.03 $\times$ 10$^{-14}$ & 1.68 $\times$ 10$^{-10}$ & 1.19 $\times$ 10$^{-11}$ & 2.50 $\times$ 10$^{-13}$ & 2.04 $\times$ 10$^{-15}$ \\
            \noalign{\smallskip}
            \noalign{\smallskip}
            \hline
         \end{tabular}
   \end{table*}

\subsection{{\tt Basic} model}
\label{Main_Basic}

The {\tt Basic} model uses the expanded network that incorporates \ce{NH2CHCO} chemistry. Figure~\ref{basic} shows the time-dependent abundances of \ce{NH2CHCO} and related products. During the collapse stage, the selected species are produced largely at early times (before 5~$\times$~10$^{5}$~yr) on the grain or ice surface and in the bulk ice. During stage 1, the gas-phase abundances of these species are very low, due to the lack of direct formation mechanisms; their presence in the gas is derived from the nonthermal desorption of surface-formed molecules. Of the species affected by the new reactions, \ce{NH2CH2CH2OH} is produced at the highest abundance, in both the gas and solid phases. Of the stable species, \ce{NH2CHCO} reaches the lowest peak abundance in both phases. 

As grain-surface production of \ce{NH2CHCO} weakens at around 5~$\times$~10$^{5}$~yr, its abundance in the outer layer of the ice falls, leading to less ejection into the gas phase. Beyond $\sim$7~$\times$~10$^{5}$~yr, its surface production picks up a little, along with the other species shown, bolstering all related species in the gas phase. Toward the end of the collapse, the gas-phase abundances of all species shown decrease to negligible values as the gas density rises, freeze-out onto the grains accelerates, and the gas-phase abundances of various species (e.g.,~C and N) involved in surface formation of \ce{NH2CHCO} become depleted.

\begin{figure*}
    \centering
    \includegraphics[width=1\textwidth]{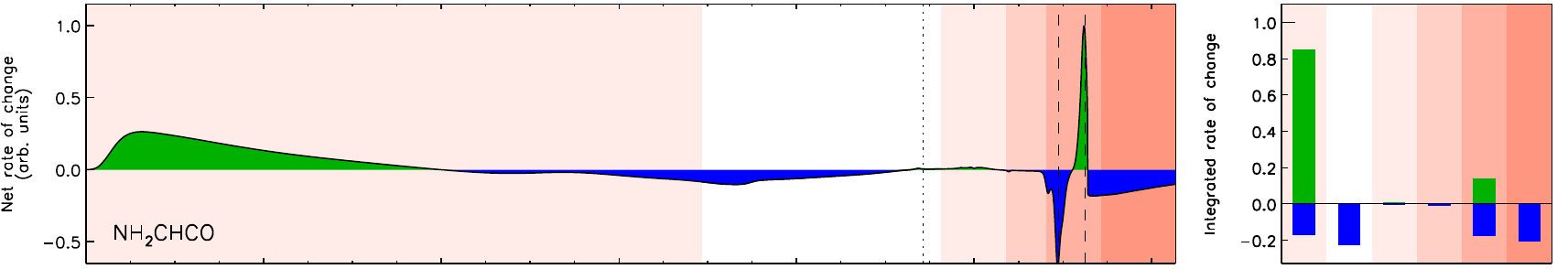}
    \includegraphics[width=1\textwidth]{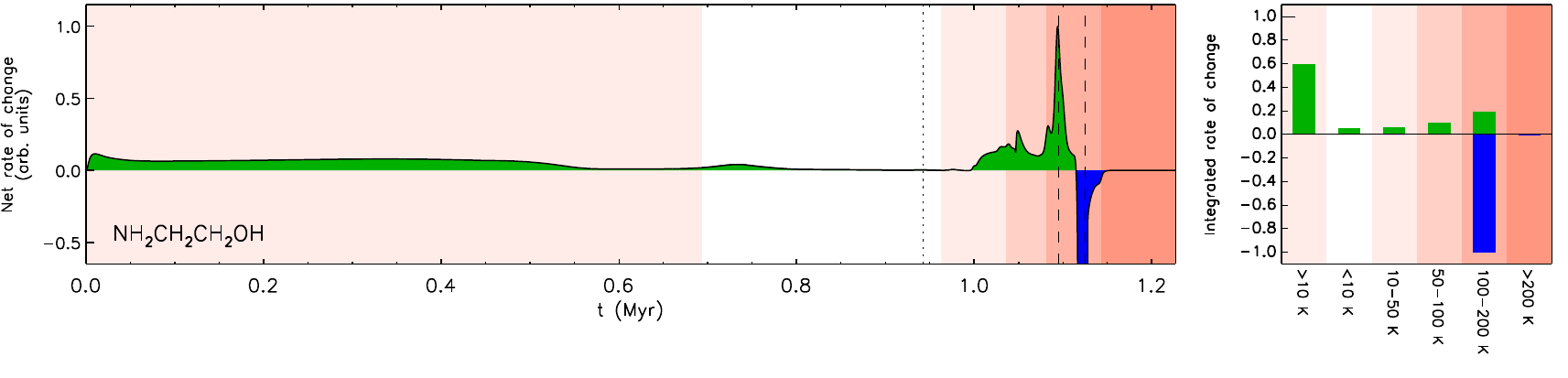}
    \caption{Left: Net rates of change in the aggregate (gas, surface and bulk phases) abundances of aminoketene (\ce{NH2CHCO}; upper) and ethanolamine (\ce{NH2CH2CH2OH}; lower) during the collapse and warm-up stages of the {{\tt Basic}} model. Net gain is shown in green, and net loss is shown in blue. The vertical dotted line indicates the start of the warm-up stage. The vertical dashed lines indicate the onset and end-point of water desorption. From left to right, the background color indicates the temperature of the dust: >10 K, 10--50 K, 50--100 K, 100--200 K, 200--400 K. Right: Integrated net rates of change over each temperature range. Positive (formation) and negative (destruction) rates are integrated independently and are both normalized to the total integrated formation rate.}
    \label{net change}
\end{figure*}

In the warm-up stage, at temperatures below $\sim$100~K, the solid-phase species are released in small quantities into the gas phase through photodesorption. As temperatures increase above $\sim$100~K, water, the main constituent of interstellar ices \citep{Boogert2015}, begins to desorb into the gas. As this occurs, species trapped in the bulk ice beneath, such as \ce{NH2CHCO} and \ce{NH2CH2CHO}, are gradually freed, allowing them also to desorb thermally. Water is lost most rapidly at $\sim$164~K, and the peak rates of thermal desorption of \ce{NH2CHCO} and \ce{NH2CH2CHO} occur very shortly after, at $\sim$173~K. The species with higher desorption energies, \ce{NH2CH2CH2OH} and \ce{NH2CH2COOH}, are retained on the grains until temperatures exceed 170 K, when they begin to thermally desorb more strongly. Once in the gas phase, all of these species are destroyed mainly by proton transfer followed by dissociative recombination. However, the rate at which that process occurs will depend on the proton affinity of the species, as discussed below.

\ce{NH2CHCO} is produced during two main periods in the model, as indicated in green in Figure~\ref{net change}, which shows the net rate of formation of aminoketene (all phases) throughout stages 1 and 2. The first period of major production comes early in the collapse stage (before 4~$\times$~10$^{5}$~yr), contributing about 85\% of total \ce{NH2CHCO} production. Because atomic C is present in high abundances in the gas phase at this time, reaction (1a) occurs almost entirely via an E-R mechanism. At temperatures above $\sim$12~K, the net positive production of \ce{NH2CHCO} is driven by a Langmuir-Hinshelwood (diffusive) reaction of \ce{CO} with \ce{CHNH2} on the grain surface.

As the temperature falls below 12~K ($\sim$5~$\times$~10$^{5}$~yr), the mobility of \ce{CO} on the surface decreases, and reaction (2) is dominated by a 3-B (nondiffusive) mechanism, whereby adsorbing C reacts with ammonia (via E-R) in the presence of CO.  Additionally, gas-phase carbon is becoming depleted at this time, resulting in less \ce{CHNH2} forming on the surface. Beyond this point, the overall production rate of \ce{NH2CHCO} becomes slightly negative until the warm-up stage. This slow loss of solid-phase aminoketene corresponds to the interconversion between \ce{NH2CHCO} and related species within the bulk ice, caused by bulk diffusion of atomic H and associated H-abstraction and addition reactions. The interconversion tends to benefit ethanolamine in particular, because, once formed, there is no barrierless, backward reaction involving atomic H that can lead to aminoketene again; the only way to produce aminoketene from ethanolamine is through photodissociation of the latter.

The decline in surface production of \ce{NH2CHCO} and related species at around $5 \times 10^5$~yr leads to a fall in corresponding gas-phase abundances in Fig.~\ref{basic}, as there becomes less and less of these species in the upper ice layer from which the gas-phase material is sourced via nonthermal desorption. There is a later uptick in gas-phase abundances beginning around $7 \times 10^5$~yr, although for \ce{NH2CHCO} this is not so apparent in Fig.~\ref{net change}. There is a net destruction of this species in the bulk ice at that time, but a small increase in its surface abundance, which through nonthermal desorption (i.e.,~photodesorption) increases the gas-phase value. For \ce{NH2CHCO}, the additional surface production is related to the nondiffusive (3-B) reactions:
\begin{gather}
    \ce{NH2CH2} + \ce{HCO} \rightarrow \ce{NH2CH2CHO} \notag \\
    \ce{NH2CH2} + \ce{CH2OH} \rightarrow \ce{NH2CH2CH2OH}, \notag
\end{gather}
which become more effective as the mobility of atomic H on the surface falls with the decreasing dust temperature, due to the longer lifetimes for surface radicals. This is followed by a sequence of H-abstraction reactions leading to \ce{NH2CHCO} production.

The second period of net \ce{NH2CHCO} production occurs in the warm-up stage, driven by the rapid loss of material from the ice surface into the gas phase by thermal desorption, allowing trapped radicals (in this case, \ce{NH2CHCHO}) in the bulk ice to become available on the surface. Abstraction by H atoms can occur either on the surface prior to desorption or in the gas phase itself. The brief spike in net production is quickly followed by net gas-phase destruction, driven by reactions with \ce{H3O+} ions, and by cosmic ray-induced photodissociation.

Ethanolamine (\ce{NH2CH2CH2OH}) is also produced most strongly during the cold collapse stage, contributing about 60\% of its total production, which is driven by successive H-addition to \ce{NH2CHCO} by diffusive H. The process proceeds primarily through the following reactions (3a, 4, 6a, and 7a):
\begin{gather}
    \label{QMCO}
    \ce{H} + \ce{NH2CHCO} \rightarrow \ce{NH2CH2CO} \notag \\
    \label{QMCHO}
    \ce{H} + \ce{NH2CH2CO} \rightarrow \ce{NH2CH2CHO} \notag \\
    \label{QMMO}
    \ce{H} + \ce{NH2CH2CHO} \rightarrow \ce{NH2CH2CH2O} \notag \\
    \label{QMMOH}
    \ce{H} + \ce{NH2CH2CH2O} \rightarrow \ce{NH2CH2CH2OH}. \notag
\end{gather}
At around $7 \times 10^5$~yr there is a small spike in net production, which is related to weak surface production of \ce{NH2CH2CH2OH} through radical addition reactions as H-atom mobility falls (see above).

The second period of major net \ce{NH2CH2CH2OH} production occurs in the warm-up stage, spread throughout the 10--100~K regime and following the same reaction scheme. Rapid destruction of \ce{NH2CH2CH2OH} occurs in the gas phase during its thermal desorption, beginning around 170~K. Once in the gas phase, it is protonated to form \ce{NH2CH2CH2OH2+}, which is then dissociated by electronic recombination. The main contributor to the protonation of \ce{NH2CH2CH2OH} is NH$_4^+$; the proton affinity of \ce{NH2CH2CH2OH} is 930.3 kJ mol$^{-1}$ (Table \ref{species}), which is much greater than that of ammonia, 853.6 kJ mol$^{-1}$ {\citep{H+L1998}}.
The ability to accept protons from \ce{NH4+} (and the inability of protonated ethanolamine to pass them on to ammonia) makes the protonation process highly efficient, leading to more rapid destruction than for species with lower proton affinities. As a result, as may be seen in Fig.~\ref{basic}, the ethanolamine fractional abundance declines rapidly compared with lower proton-affinity species and does not achieve as high an abundance in the gas phase as it does on the grain prior to desorption.
\begin{figure*}
    \centering
    \includegraphics[width=0.435\textwidth]{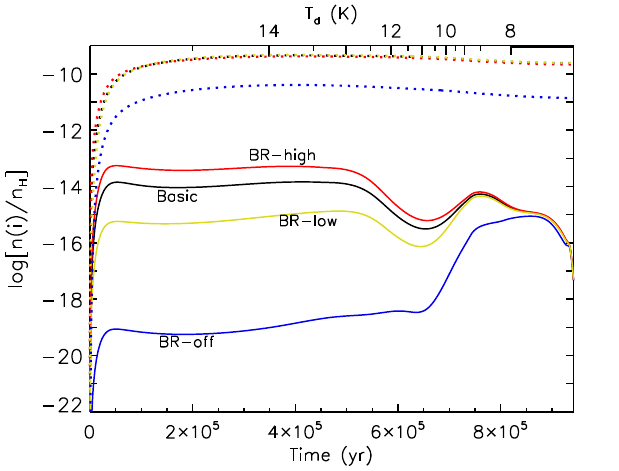}
    \includegraphics[width=0.435\textwidth]{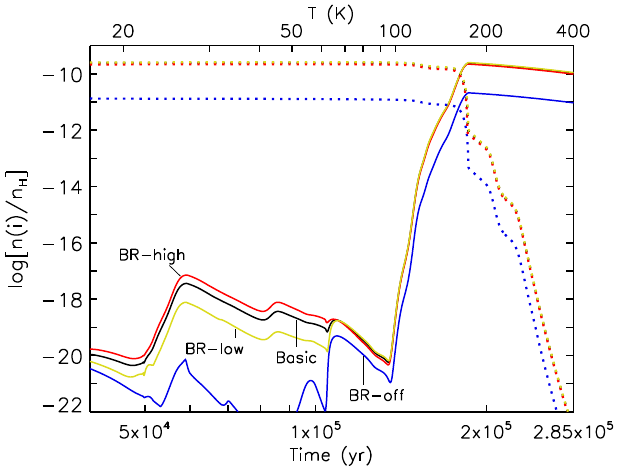}
    \caption{Fractional abundances of \ce{NH2CHCO} with respect to total hydrogen for four different setups, with the collapse stage on the left and the warm-up stage on the right. Solid lines indicate gas-phase abundances; dotted lines of the same color indicate the species on the grain (surface and bulk ice).}
    \label{multiple}
\end{figure*}
Glycine (\ce{NH2CH2COOH}) is produced in the MAGICKAL chemical network by four processes, described as reactions 5--8 in \citet{G2013}. Their reaction 8 allows \ce{NH2CH2COOH} to be produced by the addition of OH to a \ce{NH2CH2CO} radical.
The expansion of the chemical network to include \ce{NH2CHCO} chemistry leads to an increase in the abundance of \ce{NH2CH2CO} through reaction (3a). 
Comparing with \ce{NH2CH2COOH} production using only the chemical network used by \citeauthor{GH2023} (\citeyear{GH2023}; labeled here as the {\tt Old} model, with results shown in Tables~\ref{all_st1_output} and~\ref{all_st2_output}), the reaction rate between the \ce{NH2CH2CO} radical and \ce{OH} increases by $\sim$30\%, which drives the overall production of \ce{NH2CH2COOH} up by factor of two in the {\tt Basic} model. Like \ce{NH2CH2CH2OH}, \ce{NH2CH2COOH} is destroyed rapidly in the warm-up stage by protonation from \ce{NH4+} because of its relatively high proton affinity of 886.5 kJ mol$^{-1}$ {\citep{H+L1998}}. For a more in-depth description of \ce{NH2CH2COOH} chemistry in the collapse and warm-up stages, see \citet{G2013} and \citet{GH2023}. 

\subsection{Branching ratio-dependent models}
\label{CHNH2-dependent}

With the addition of \ce{NH2CHCO} chemistry to the network, one goal of this work is to determine the dominant mechanism of formation for \ce{NH2CHCO} and its products. As may be seen in Table~\ref{reactions}, aminoketene can be formed in our network either via reactions (1a) and (2), or by H-abstraction from, for example, ethanolamine (which can be formed by radical addition, not just hydrogenation of \ce{NH2CHCO}). 
The models {\tt BR--high}, {\tt BR--low} and {\tt BR--off} have been run to determine how the efficiencies of reactions (1a,b) and (2) affect the production of \ce{NH2CHCO} and its products. In the {\tt BR--high} and {\tt BR--low} models, either the \ce{CHNH2} or \ce{CH2NH} product was favored, as shown in Table~\ref{input}. In the {\tt BR--off} model, the efficiencies of reactions (1a,b) and (2) were set to zero to prevent them from occurring at all.

The abundance of \ce{NH2CHCO} over time is shown across the four models in Figure~\ref{multiple}. The qualitative behavior of the collapse and warm-up chemistry in both the {\tt BR--high} and {\tt BR--low} setups is almost identical to the behavior in the {\tt Basic} setup, but they do produce a modest change in the abundances shown in Tables~\ref{all_st1_output} and~\ref{all_st2_output}.

During the collapse phase, in the case of the {\tt BR--high} setup, the final solid-phase abundance of ethanolamine is increased by $\sim$0.25 orders of magnitude when compared to the {\tt Basic} setup. In the {\tt BR--low} setup, it is decreased by a similar degree. The abundance of \ce{NH2CH2CH2OH} could therefore plausibly vary by $\sim$0.5 orders of magnitude, depending on the true branching ratio of the reaction. Similar ratios are preserved in the peak gas-phase abundance of \ce{NH2CH2CH2OH} during the warm-up stage in each model, albeit at overall much lower values, due to the efficient gas-phase destruction described in Sec.~\ref{Main_Basic}.

The abundances of the other species in the models in general are yet less sensitive to the choice of branching ratio; the final solid-phase abundances (collapse stage) and peak gas-phase abundances (warm-up stage) of those molecules are essentially unchanged between models, except in the case of \ce{NH2CHCO}. Indeed, the latter species demonstrates a marginally higher overall solid-phase abundance during the collapse stage of the {\tt BR-low} model, as compared with the {\tt Basic} or {\tt BR-high} setups. Meanwhile the gas-phase abundances of \ce{NH2CHCO} at early times span approximately two orders of magnitude between the models, this time in favor of the {\tt BR-high} model. Again, all these gas-phase abundances are still very low and are reflective of nonthermal desorption from the outer dust-grain ice layer.

This behavior during the collapse stage -- in which the gas-phase and ice-surface abundances can be somewhat enhanced, while the bulk-ice abundances are fairly stable -- is clear evidence that the balance between \ce{NH2CHCO}, \ce{NH2CH2CHO}, \ce{NH2CH2CH2OH}, and all the intermediate radicals continues to evolve once all of these species have become incorporated into the bulk ice. Atomic H, which is allowed to diffuse within the bulk ice following its production via photodissociation of various major ice component molecules, can attack each of these species, ultimately leading to an equilibration of abundances that is reflected consistently across each of the {\tt Basic}, {\tt BR-low} and {\tt BR-high} models.

For the {\tt BR--off} setup, \ce{NH2CHCO} production is significantly decreased, but it is not zero because other formation pathways exist that are not dependent on reactions (1) and (2). These include the radical-addition reactions forming \ce{NH2CH2CHO} and \ce{NH2CH2CH2OH}, which are then followed by H-abstraction, as noted in Sec.~\ref{Main_Basic}.

In summary, the low gas-phase abundances obtained under low-temperature conditions are much more varied between models than the ice abundances, while the gas-phase abundances at high temperatures, post-desorption, are naturally reflective of the more stable bulk-ice abundances. Thus, the gas-phase abundances that would be representative of a hot core are essentially indistinguishable between the {\tt Basic}, {\tt BR--high} and {\tt BR--low} models. While reactions (1a) and (2) are essential to form \ce{NH2CHCO} and its related products, the precise branching ratio between reactions (1a) and (1b) is tempered by the interconversion of \ce{NH2CHCO}, \ce{NH2CH2CHO} and \ce{NH2CH2CH2OH} within the bulk ice, via reactions with atomic H.

\subsection{Shock models}\label{results:shock}

\begin{figure}
    \centering
    \includegraphics[width=0.435\textwidth]{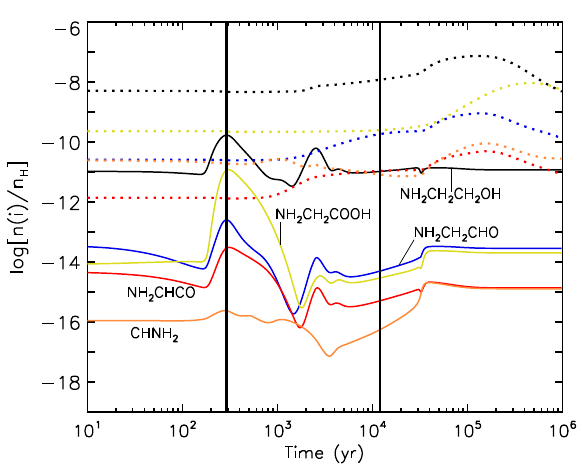}
    \caption{Fractional gas-phase abundances of \ce{NH2CHCO} and relevant products with respect to total hydrogen. The results correspond to the {\tt S--VHCR} setup, with the collapse stage shown on the left and the shocked stage shown on the right. Solid lines indicate gas-phase abundances; dotted lines of the same color indicate the species on the grain (surface and bulk ice).}
    \label{VHCR}
\end{figure}

Finally, the four shocked models, {\tt S--LCR}, {\tt S--MCR}, {\tt S--HCR}, and {\tt S--VHCR}, are included to understand the production of \ce{NH2CHCO} and its products in shocked regions, since \ce{NH2CH2CH2OH} was detected in the shocked molecular cloud G+0.693. Figure~\ref{VHCR} shows the time-dependent abundance of \ce{NH2CHCO} and some of its products for the shock model with CRIR $\zeta = 1000 \zeta_0$ ({\tt S--VHCR}), which most closely models the conditions of G+0.693. Because of a lower final density than in the hot-core models, the visual extinction only reaches $\sim$11 mag and the temperature, which falls as a function of the visual extinction, reaches 8.6~K at the end of the collapse stage.

In the shock models, the newly added species follow the same production mechanisms as in the collapse stage of the {\tt Basic} model, with \ce{NH2CHCO} still forming through an E-R reaction between C and \ce{NH3}, followed immediately by reaction with CO. The other species of interest are also still formed by the previously discussed mechanisms, with \ce{NH2CH2CHO} and \ce{NH2CH2CH2OH} being produced by the H-addition reactions (3a, 4, 6a, and 7a). 

The gas-phase chemical abundances begin to increase rapidly just before 200~yr, as seen in Figure~\ref{VHCR}. This increase is a direct result of the shock-induced sputtering of the icy grains.  
Importantly, this process allows heavier species, such as those we focus on in this paper, to be released into the gas phase at a much higher rate than they would if desorption were driven only by low-temperature thermal, reactive, or photon-induced desorption. The low peak dust temperature reached in the shock means that thermal desorption remains unimportant for all but very volatile species.

Most of the molecules reach their peak abundances shortly after the shock begins, at around 300 years into the model (indicated by the first vertical line in Figure~\ref{VHCR}). The precise peak values are indicated in Table~\ref{all_st2_output}. Once in the gas phase, these species are destroyed quickly by ion-molecule reactions with abundant ions (C$^+$ and H$_3^+$ in particular) and by cosmic ray-induced photodissociation associated with the high CRIR. The peak abundances achieved for each species are dependent not only on the sputtering rates, but also on the rates of gas-phase destruction, which in turn are dependent on the dipole moments of each species, as well as the overall gas density.

It is notable that the ice mantles are not completely desorbed during the shock; indeed, only around 5\% of the ice is desorbed at the 300~yr mark. By around 400~yr, sputtering is no longer the main desorption mechanism for most species, due to the brief period when $v_{\mathrm{dust}}-v_{\mathrm{neut}}$ is large, and the relatively high rate of photodesorption in the {\tt S--VHCR} model.

Importantly, due to the comparatively low gas-phase abundance of ammonia in the post-shock gas (versus the post-desorption abundance in hot core models), the enhanced destruction of species with high proton affinities by reactions with NH$_4^+$ is diminished. Glycine and ethanolamine are therefore not destroyed at a substantially greater rate than the other large N-bearing species. Glycine also shows an especially strong increase compared with its pre-shock abundance; the latter is determined by the availability of glycine in the surface layer of the ice prior to the passage of the shock, which is relatively low compared with, for example, ethanolamine. 

Beyond the 300~yr mark, the gas density continues to rise, driving gas-phase ion-molecule destruction more rapidly, which leads to lower abundances of the plotted species than prior to the shock.
There is a second increase in abundance for those species (excluding CHNH$_2$) that begins just before $2 \times 10^{3}$~yr; this is driven by increased concentrations of the various species in the surface layer of the ice in particular. That grain-surface rise is related to the cooling of the dust back to its minimum temperature, which is complete by around $2.5 \times 10^{3}$~yr. The drop in dust temperature reduces the speed of H-atom recombination with surface radicals, increasing the probability of nondiffusive radical-radical reactions that lead to COM production.

After the second peak, the bulk-ice abundances of various molecules (including major species like water) continue to grow, as material that was sputtered or otherwise desorbed is re-accreted. This includes CO, as well as its ion-molecule destruction products, atomic C and O. The re-accretion of these species leads to new grain-surface chemistry that can produce \ce{CHNH2}, and hence COMs such as ethanolamine.

The second vertical line in Figure~\ref{VHCR} indicates the time at which the gas temperature has cooled down to 100~K, the initial temperature of the gas before the shock. After this time, the gas-phase abundances remain relatively stable, while grain-surface ice abundances grow due to re-accretion, as described above. Toward the end of the model, due to the extreme CRIR, photodissociation of the ices reduces the abundances of the larger COMs, resulting in a net production of simpler species.

\begin{figure*}[h!]
\begin{minipage}[t]{0.435\textwidth}\vspace{0pt}
\includegraphics[width=\textwidth,height=0.25\textheight]{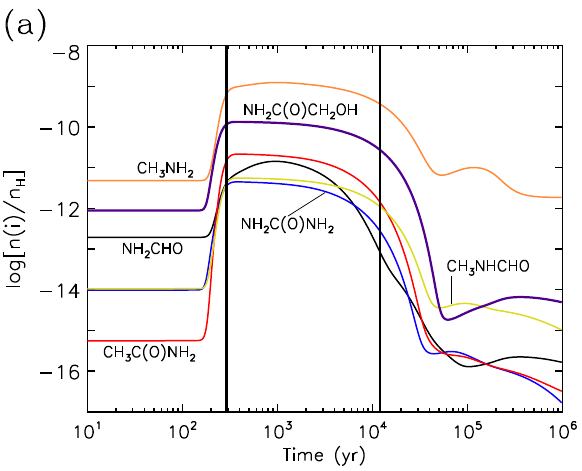}
\vfill 
\includegraphics[width=\textwidth,height=0.25\textheight]{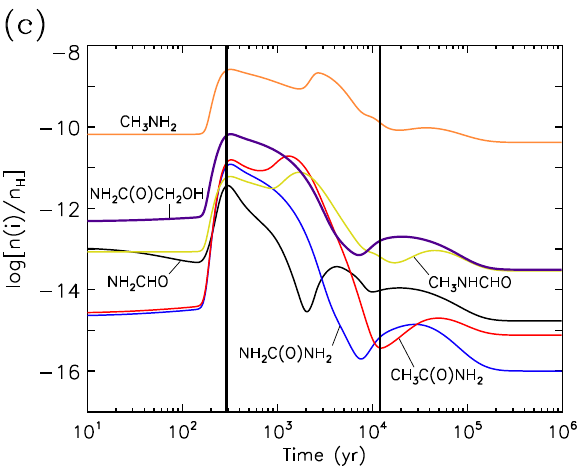}
\end{minipage}\hspace{0.065\textwidth}
\begin{minipage}[t]{0.435\textwidth}\vspace{0pt} 
\centering 
\includegraphics[width=\textwidth,height=0.25\textheight]{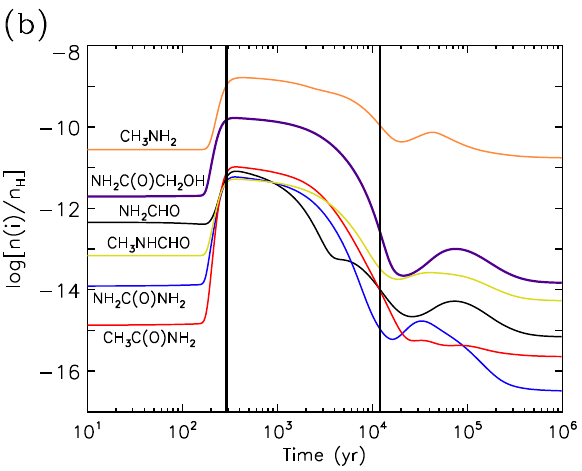}
\vfill 
\includegraphics[width=\textwidth,height=0.25\textheight]{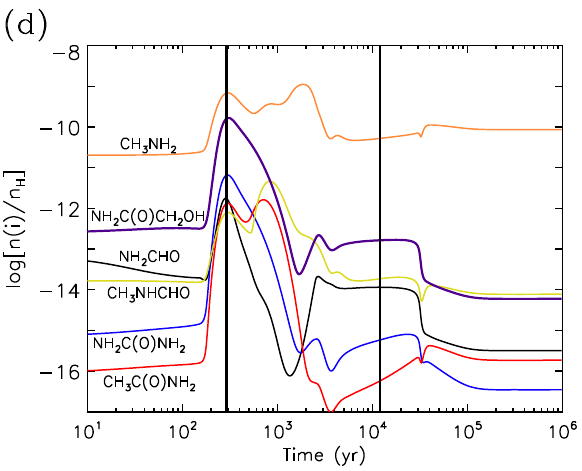} 
\end{minipage}
\caption{Fractional gas-phase abundances with respect to total hydrogen of nitrogen-bearing species observed in G+0.693. The first vertical line indicates the approximate time for the peak abundances of the species. The second vertical line indicates the time when the gas temperature has returned to the starting temperature of 100~K. The four panels represent the four setups with increasing CRIR (a) {\tt S--LCR}, {\tt S--MCR}, {\tt S--HCR}, and {\tt S--VHCR}.}
\label{mag_obs}
\end{figure*}

The abundances with respect to total hydrogen of several nitrogen-bearing species that have been observed in G+0.693 are shown in Figure~\ref{mag_obs}, for all four shocked models. The most notable trend in these models is that with lower CRIR values, the gas-phase lifetime of the molecules during the shock increases. Higher CRIR values increase the rate of gas-phase destruction through ion-molecule reactions and enhanced cosmic ray-induced photodissociation. 

Without high cosmic-ray ionization rates, the main loss mechanism of the post-shock gas-phase species is their re-accretion onto the grains, without first being broken down. Higher CRIR values lead to a much more complex behavior for all species during the passage of the shock. 

At late times in all of the shock models, the abundances of most of the COMs continue to fall to values lower than those before the shock. This caused by the elevated gas density, which allows more rapid accretion onto the grains.

It is also notable that, before the passage of the shock, the gas-phase abundances of the plotted complex species, while rather low, generally take higher values in the {\tt S--MCR} and {\tt S--HCR} models than in the {\tt S--LCR} and {\tt S--VHCR}. This is due to the more rapid photodesorption of these species, caused by the CR-induced UV field, as CRIR is increased. In the {\tt S--VHCR} model, this effect is balanced by the more rapid gas-phase destruction of those species by various cosmic ray-related processes.

\section{Discussion}
\label{disc}

Given the many uncertainties regarding the production mechanisms of prebiotic molecules in the ISM, calculating the abundance of \ce{NH2CHCO} on dust grains and in the gas of the interstellar medium may prove vital to our understanding of the abundances of other amine-bearing species. Our modeling results predict a peak solid-phase abundance of 2.45~$\times$~10$^{-10}$~n$_{\mathrm{H}}$ for aminoketene (\ce{NH2CHCO}) in the warm-up stage of the {\tt Basic} model, but most of it is formed at low temperatures. The idea that COMs are formed on cold grain surfaces is supported by their detections in cold ($\sim$10~K) prestellar cores and molecular clouds (\cite{JS2016}; \citet{Zeng2018}; \citet{Riv2018}). In our models, reactions (1a) and (2) are the primary mechanism by which \ce{NH2CHCO} is formed. In fact, the aminoketene produced in the model is efficiently converted onward to increase the abundances of larger species, including ethanolamine in particular.

The model indicates also that reaction (1a) will proceed via an E-R process as gaseous C adsorbs onto \ce{NH3} formed on the surface by the hydrogenation of atomic N. Under the translucent initial conditions assumed in the model, most of the singly ionized carbon in the gas that would originate at lower densities ($<$1000 cm$^{-3}$) and visual extinctions is soon enough converted into neutral form, with some being incorporated into carbon monoxide \citep[see, e.g.,][]{Snow2006}. Therefore, the abundance of C in the gas phase is high enough that the E-R formation of \ce{CHNH2} can be substantial, especially during the early evolution of a cold core. Under these conditions, our modeled \ce{NH3} ice has a fractional abundance of $\sim$7\% with respect to \ce{H2O}, in agreement with typical observational values \citep{Oberg2016}. At the early times ($< 5\times 10^5$~yr) when atomic C is most abundant in the gas, the calculated visual extinction is low, and thus dust temperatures are higher ($>$12~K), making CO fairly mobile. The subsequent reaction of newly formed \ce{CHNH2} with CO therefore typically occurs through a diffusive reaction. The resulting aminoketene can be further hydrogenated to glycinal and ethanolamine by surface H atoms. However, as new ice layers form, and all of these species become incorporated into the bulk ice, they can be further interconverted by diffusive H within the bulk. This process appears to have a strong effect in determining the final relative ratios of these species.
Finally, solid-phase \ce{NH2CHCO} desorbs efficiently into the gas phase once temperatures are high enough ($\sim$170 K). Desorption of water is more or less complete by the time aminoketene desorption becomes efficient.
While barely detectable amounts of gas-phase ethanolamine (\ce{NH2CH2CH2OH}) might be present during the cold core stage (around $10^{-11} n_{\mathrm{H}}$), dependent on the efficiency of photodesorption, none of the other molecules introduced in this study, including aminoketene, should reach such abundances until the dust-grain ices have sublimated. 

Due to its high reactivity, the slow thermal desorption of aminoketene during the hot period of the hot core could lead to its efficient polymerization on the icy grain surfaces, and the formation of various larger prebiotic COMs including peptides \citep{Krasnokutski2024, Suhasaria25}. This could lead to a lower abundance of this molecule in the gas phase than predicted by the model, which does not include any reaction between aminoketene molecules. Based on an analysis of the model results, it appears likely that thermal desorption of aminoketene would be competitive with such a reaction at the temperatures at which aminoketene becomes both mobile and relatively abundant on the surface, leading to a strong preference for desorption over the production of larger species. However, re-adsorption of gas-phase aminoketene at temperatures above its desorption temperature would provide many further opportunities for aminoketene molecules to react with each other. Determining the overall yield of larger aminoketene products, including peptides, during the hot stage thus requires an explicit model. We leave this effort to future work.

The solid-phase detection of complex nitrogen-bearing interstellar molecules is limited, in spite of the recent identification of oxygen-bearing COMs using JWST \citep{Rocha2024}. For example, a recent estimate of the upper limit of ethanolamine abundance in the solid state provided a value of $\sim$15\% with respect to H$_2$O \citep{Suhasaria24}. This high value is obtained because the 1607 cm$^{-1}$ absorption band used to estimate its abundance could also originate from many other molecules. Therefore, a more useful metric for understanding \ce{NH2CHCO} chemistry may be its abundance in the gas. Due to its formation on grains, we do not expect it to be detectable in the gas phase at low temperatures. However, in our {\tt Basic} hot-core model, the peak, post-sublimation gas-phase abundance is 2.40~$\times$~10$^{-10}$~n$_{\mathrm{H}}$, essentially identical to the peak solid-phase abundance. This indicates that this molecule is efficiently desorbed without undergoing substantial surface or gas-phase destruction that would limit its peak abundance. Provided rotational spectroscopic data for this species, it should in principle be detectable. An observational estimate of the abundance of aminoketene in the gas phase could greatly improve the accuracy of this part of the chemical models.

Gas-phase detection of aminoketene would require knowledge of its rotational spectrum, but due to the molecule's high reactivity, it may be difficult to generate sufficient concentrations in the gas phase for laboratory measurements. An alternative is to use calculated rotational spectra for comparison with observations \citet{Alberton2024}. However, the insufficient precision of these kinds of calculations often poses a challenge to the detection of molecules with low abundances.

\begin{figure}
    \centering
    \includegraphics[width=0.48\textwidth]{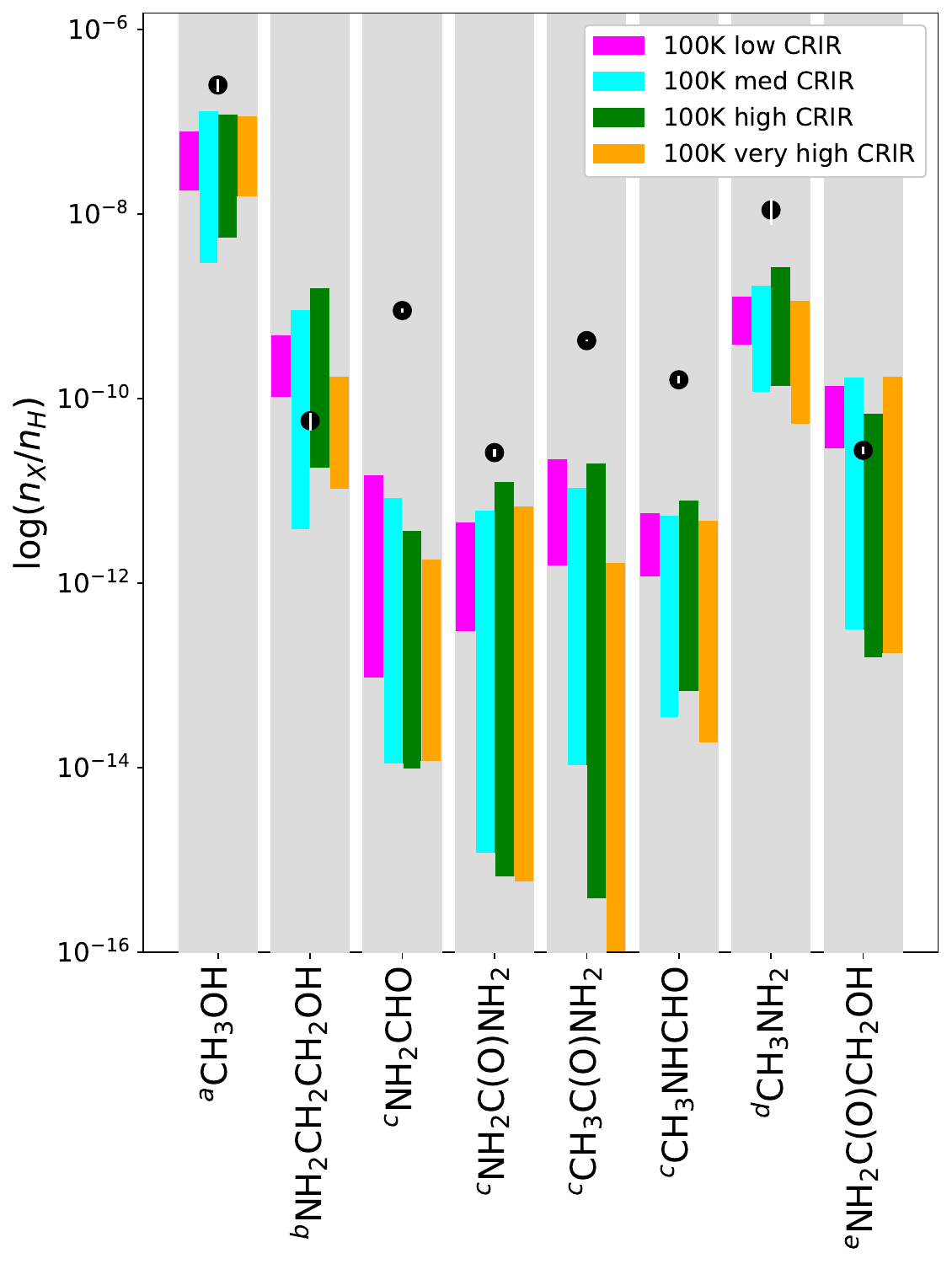}
    \caption{Comparison of observed fractional abundances with respect to \ce{H2} of a selection of COMs that have been observed in G+0.693 to their predicted gas-phase abundances for the {\tt S--LCR}, {\tt S--MCR}, {\tt S--HCR}, and {\tt S--VHCR} setups used in this work. Observational values correspond to column density ratios. The maximum values included correspond to the peak abundance in the shocked stage. The minimum values included correspond to when the gas temperature has cooled to 100 K, an estimate of the gas kinetic temperature according to \citet{Zeng2018}. The observation values are represented as black dots, with errors included as white vertical lines. The values are taken from multiple sources a) \citet{Torres2008}, b) \citet{Riv2021}, c) \citet{Zeng2023}, d) \citet{Zeng2018}, and e) \citet{Riv2023}.}
    \label{obs vs. mag}
\end{figure}

Although aminoketene has not yet been detected in the ISM, ethanolamine has been identified with a reported abundance of (0.9--1.4)~$\times$~10$^{-10}$~n$_{\mathrm{H_2}}$ in the molecular cloud G+0.693 \citep{Riv2021}. This quiescent cloud is located near the active star-forming region Sgr B2(N) of the central molecular zone of the Milky Way \citep{Zeng2018}. Other amine-bearing species have been detected in this cloud, including formamide (\ce{NH2CHO}), urea (\ce{NH2C(O)NH2}), acetamide (\ce{CH3C(O)NH2}), and glycolamide (\ce{NH2C(O)CH2OH}) \citep{Zeng2018, Zeng2023, Riv2023}. As a means of determining the overall accuracy of the shock models, we can compare the simulated abundances of these amine-bearing species with their observed values. Figure~\ref{obs vs. mag} depicts this comparison; predicted abundances from all four of the shocked models are included, based on the range of values obtained for each molecule between the peak gas-phase abundance and the moment in the model when the gas temperature returns to its pre-shock value (100~K). Values for methanol (\ce{CH3OH}) are also shown for reference. 

Of the species shown in Figure~\ref{obs vs. mag}, \ce{NH2CH2CH2OH} and \ce{NH2C(O)CH2OH} have observed abundances that fall within the modeled abundances for three of the four simulations. Additionally, the observed abundances of \ce{CH3OH}, \ce{NH2C(O)NH2}, and \ce{CH3NH2} are within a factor of two higher than the modeled abundances. The abundances of \ce{NH2CHO}, \ce{CH3C(O)NH2}, and \ce{CH3NHCHO} in the models differ from their observational values by about one to two orders of magnitude. \ce{NH2CH2CH2OH} is the only species that is overproduced in the shock models. However, the overall relative trend for each of these molecules is good, with \ce{NH2CHO} appearing to be the furthest outlier.

This systematic shortfall of the models versus the observations would indicate, perhaps, either that the observational determination of the H$_2$ column density is an underestimate, or that the chemical models have too little sputtering. The latter option is certainly possible; as noted in Sec.~\ref{results:shock}, only around 5\% of the ices are ejected during the shock, so a greater rate of sputtering or a more sustained period of ejection could increase the gas-phase values. The assumed shock speed of 20~km/s would have to be substantially increased in the model to achieve an order of magnitude greater sputtering loss, but an increase in the assumed dust grain size from 0.1~$\mu$m to 1~$\mu$m would raise the peak $v_{\mathrm{dust}}-v_{\mathrm{neut}}$ value from $\sim$2.06 to 5.21~km/s, while maintaining $v_{\mathrm{dust}}-v_{\mathrm{neut}}$ above 1~km/s for around 4 times longer. The dust grain size is therefore a major determinant in the degree of ice lost to the gas. We note that the dust temperature would still be too low to produce thermal desorption of heavy species.

On the assumption that the relative gas-phase abundances produced in the model are reliable, glycine would be expected to have an abundance around an order of magnitude lower than ethanolamine in G+0.693. The abundance of glycinal, which does not presently have an available spectrum for its detection in the gas phase, is strongly dependent on the CRIR, but under more typical galactic ionization rates might have a gas-phase abundance similar to glycine. Aminoketene is unlikely to be detectable.

The individual abundance of ethanolamine is in fact a good match to the observed value, but could be considered overproduced in the context of a systematic underestimation of fractional abundances. The abundance of ethanolamine in the modeled ices is dependent to a large degree on the balance of interconversion between glycinal and its pre-cursor aminoketene, and the conversion of glycinal onward to ethanolamine. In our network, and based on our activation energy-barrier calculations, only photodissociation is capable of reversing this process to backward convert ethanolamine toward aminoketene. The overproduction of \ce{NH2CH2CH2OH} could therefore be indicative of an insufficiently effective or missing mechanism of backward conversion. 

Alternatively, one may consider that the shock model, following past treatments, assumes a uniform sputtering rate for all species that varies only according to the abundance of each species in the ice. Sputtering rates might more accurately be based on the binding energies of the individual species, rather than a generic value representative of water. This could have the effect of decreasing the sputtering rate of ethanolamine, while also increasing, for example, the rate for formamide (\ce{NH2CHO}). We also note that our network includes the reaction of NH$_2$ with H$_2$CO to produce formamide at the rate determined by \citet{Vazart2016}, but it does not contribute significantly to the production of this molecule in the shock model.

The shock models also indicate two other findings; firstly, although the peak abundances of COMs achieved using various CRIR values from $\zeta = 1.3 \times 10^{-17}$ -- $1.3 \times 10^{-14}$~s$^{-1}$ do not vary wildly, the speed of destruction following the achievement of the peak value does vary. Destruction of COMs is around ten times fast for a CRIR value $\zeta = 10 \, \zeta_0$. Secondly, the ion-molecule destruction of these species is driven by charge-exchange reactions with the C$^+$ ion, and not protonated from small molecular ions. As a result, the relative proton affinities (PAs) of the COMs are unimportant to their destruction rates, and species with large PAs are therefore not disproportionately destroyed \citep{GH2023}. Thus, shocked regions may indeed be a better target (versus hot cores) for the detection of amine-bearing COMs that are likely to have large PA values, including glycine and ethanolamine.

In both the shock models and the hot core models, the gas phase abundances of the species explored here are strongly affected by the rates of UV-induced photodesorption at low temperatures. A blanket quantum yield of $10^{-3}$ is assumed in our models for new species. However, the experimental evidence for efficient photodesorption of COMs is sparse \citep{MartinD2016, Bertin2016}. Nevertheless, the most important findings of this study are not influenced by this uncertainty. We generally find that none of the COMs explored here would be abundant enough in the gas-phase to be easily observable in cold cores, while the peak abundances in the shock models are still reflective of the rate of sputtering and not photodesorption. We note also that further investigation of the barrier-mediated reactions included in our network, at higher levels of theory, would be helpful in producing the most accurate network possible.
We hope to combine such methods in future with a more explicit calculation of tunneling transmission rates using the calculated potential profile itself, rather than approximation to a rectangular barrier.

\section{Conclusions}
\label{concl}
The inclusion of \ce{NH2CHCO} and \ce{NH2CH2CH2OH} chemistry in the chemical network of MAGICKAL has provided an estimate for the solid-phase and gas-phase abundances of the species, along with a slightly increased value for the abundance of \ce{NH2CH2COOH}. The results support the theory that \ce{NH2CHCO} can be formed efficiently on interstellar dust grains by barrierless reactions of C, \ce{NH3}, and CO. The efficient formation of \ce{NH2CH2COOH} and \ce{NH2CH2CH2OH} are also triggered by reactions (1a) and (2). 

Additionally, the application of a shock treatment provides estimates of the abundances of nitrogen-bearing species that have been observed in the G+0.693-0.027 molecular cloud. Based on the comparison of results from the shock treatments with observational values for those species, our models show good agreement and are consistent with previous findings that the CRIR of this cloud is likely 100 to 1000 times greater than the canonical Galactic rate.

Although the predicted abundances of the affected amine-bearing species are on the lower end of plausible detectability, their detection is possible. Given the detection of ethanolamine in a shocked region \citep{Riv2021}, along with our results indicating that glycine and ethanolamine are not destroyed by \ce{NH4+} as substantially in a shocked region compared to a hot core, shocked regions similar to G+0.693 remain the most likely place to detect these species.

The CRIR may also play an important role in determining the likelihood of detecting these species. A lower CRIR means longer gas-phase lifetimes after the passage of a shock, while a higher CRIR limits the lifetime and hence the time frame of detectability. Although the exact timescale of the shocks is unknown, further constraints on the values of CRIR in G+0.693 and other shocked regions will reveal the regions in which these species are most likely to survive and therefore be detected.

The main conclusions from this work are summarized below:
\begin{enumerate}
  \label{list_conclusions}
  \item Our models indicate that aminoketene (\ce{NH2CHCO}) is formed primarily on interstellar dust grains by barrierless reactions of C, \ce{NH3}, and CO. These reactions are not energy dependent, so they are efficient at low temperatures, such as those experienced in the cold collapse stage of a molecular cloud.
  \item Considering that aminoketene could be efficiently desorbed in hot cores at high temperatures without significant destruction, this would result in a predicted abundance of 2.40~$\times$~10$^{-10}$~n$_{\mathrm{H}}$, which means that its detection in hot cores is plausible but would require a well-defined rotational spectrum.
  \item Ethanolamine (\ce{NH2CH2CH2OH}) is formed primarily by H-addition to aminoketene, but it can also be produced through other pathways, such as nondiffusive radical-radical reactions. Similarly, glycine may also be formed from aminoketene or from nondiffusive radical-radical reactions.
  \item Shocked regions may be the best place to search for amine-bearing species with proton affinities greater than 853.6~kJ~mol$^{-1}$, given that they are less likely to be destroyed by \ce{NH4+} in these regions; this is due to lower gas densities and (possibly) the incomplete desorption of NH$_3$ along with the other ices. This theory is supported by the detection of multiple amine-bearing species in G+0.693 compared to other regions.
  \item Interconversion between aminoketene and ethanolamine, through H-addition, leads to an equilibration of their abundances in the bulk ice, despite varying conditions affecting the surface abundances. Therefore, early gas-phase abundances at lower temperatures are reflective of the conditions affecting surface chemistry, while late gas-phase abundances at high temperatures reflect the stable abundances in the bulk ice.
\end{enumerate}

This work could have important astrochemical implications, as these barrierless reactions with atomic carbon could be crucial for the formation of various prebiotic molecules, including amino acids and aminoketene.  Current models do not include the polymerization reactions because the energy barriers for these reactions are not known. In addition, the polymerization of aminoketene on the surface of individual dust grains would not appear to be very efficient, given the solid-phase abundances in the models. However, organic molecules formed via atomic carbon chemistry on dust at low temperatures could be the building blocks of comets and asteroids, where liquid chemistry could provide very favorable conditions for aminoketene polymerization, leading to peptides. These celestial bodies may also have been responsible for delivering such pristine organic material to Earth and exoplanets.

\begin{acknowledgements}
    We thank V. Rivilla and M. Sanz-Novo for very helpful discussions of the conditions in G+0.693-0.027.
    We thank S. Viti for advice regarding the use of the shock modules in UCL\_CHEM.
    S.A.W. and R.T.G. thank the National Science Foundation for funding through the Astronomy \& Astrophysics program (grant number 2206516). S.A.K is grateful to DFG (grant number 413610339). 
\end{acknowledgements}
   \bibliographystyle{aa}
   \bibliography{willis}

\end{document}